%% file: main.tex
\documentclass[conference, 9pt]{IEEEtran}
\usepackage{caption}
\usepackage{subcaption}
\usepackage{nopageno}
\usepackage{graphbox} % it depends on "graphicx" package
\usepackage{amsmath,amsfonts,amsbsy,latexsym}
\usepackage{multirow}
\usepackage{makecell}
\usepackage{xcolor}
\usepackage{colortbl}
\usepackage{graphicx}
\usepackage{tabulary}
\usepackage{url}
\usepackage{booktabs}
\usepackage{cite}

\include{packages}
\begin{document}

% \title{\Huge{Inherently-Parallel Differential Sampling of Multi-Level Digital Circuits}}

\title{{\sc Demotic}: A \underbar{D}iffer\underbar{e}ntiable Sa\underbar{m}pler \\ f\underbar{o}r Mul\underbar{t}i-Level D\underbar{i}gital \underbar{C}ircuits}

% \title{Conference Paper Title*\\
% \author{\IEEEauthorblockN{
%     Arash Ardakani, 
%     Minwoo Kang,
%     Kevin He,
%     Qijing Huang,
%     Vighnesh Iyer,
%     Suhong Moon,
%     John Wawrzynek
% }\vspace{-0.75cm}
% }
\bstctlcite{IEEEexample:BSTcontrol}
\makeatletter
    \newcommand{\linebreakand}{%
      \end{@IEEEauthorhalign}
      \hfill\mbox{}\par
      \mbox{}\hfill\begin{@IEEEauthorhalign}
    }
    \makeatother

\author{
    \IEEEauthorblockN{Arash Ardakani}
    \IEEEauthorblockA{
    \textit{University of California, Berkeley}\\
    arash.ardakani@berkeley.edu}
    \and
    \IEEEauthorblockN{Minwoo Kang}
    \IEEEauthorblockA{
    \textit{University of California, Berkeley}\\
    minwoo\_kang@berkeley.edu}
    \and
    \IEEEauthorblockN{Kevin He}
    \IEEEauthorblockA{
    \textit{University of California, Berkeley}\\
    kevinjhe@berkeley.edu}
    \and
    \IEEEauthorblockN{Qijing Huang}
    \IEEEauthorblockA{
    \textit{NVIDIA}\\
    jennyhuang@nvidia.com}
    \linebreakand
    \IEEEauthorblockN{Vighnesh Iyer}
    \IEEEauthorblockA{
    \textit{University of California, Berkeley}\\
    vighnesh.iyer@berkeley.edu}
    \and
    \IEEEauthorblockN{Suhong Moon}
    \IEEEauthorblockA{
    \textit{University of California, Berkeley}\\
    suhong.moon@berkeley.edu}
    \and
    \IEEEauthorblockN{John Wawrzynek}
    \IEEEauthorblockA{
    \textit{University of California, Berkeley}\\
    johnw@berkeley.edu}
}

% \author{\IEEEauthorblockN{
%     Arash Ardakani, 
%     Minwoo Kang,
%     Kevin He,
%     Qijing Huang,
%     Vighnesh Iyer,
%     Suhong Moon,
%     John Wawrzynek
%     \thanks{\textsuperscript{*}Both authors contributed equally.}
% }\\
% \IEEEauthorblockA{University of California, Berkeley\\
% \{arash.ardakani, minwoo\_kang\}@berkeley.edu
% }
% }
%%%%5%%%%%%%%%%%%%%%%%%%%%%%%%%%%%%%%%%%%%%%%%%%%%%%%%%%%

\newcommand\blfootnote[1]{%
  \begingroup
  \renewcommand\thefootnote{}\footnote{#1}%
  \addtocounter{footnote}{-1}%
  \endgroup
}

\definecolor{main}{HTML}{4472C4}    % setting main color to be used
\definecolor{sub}{HTML}{EBF4FF}     % setting sub color to be used
\newcommand{\com}[1]{{\color{red}\sf{[#1]}}}
\newcommand{\OURS}{{\sc Demotic}}
%%%%%%%%%%%%%%%%%%%%%%%%%%%%%%%%%%%%%%%%%%%%%%%%%%%%%%%%

% \input{_0_abstract}

\maketitle
\input{00_abstract}
\input{01_introduction}

\input{04_preliminaries}

\input{03_methodology}

\input{05_experiments}

\input{02_background}

\input{06_conclusion}
% \bibliography{references, references_scheduling, references_accel, references_algo}

\bibliographystyle{IEEEtran}
\bibliography{sat_sampling.bib}

\end{document}

%% file: packages.tex
\usepackage{smartdiagram}

\usepackage[activate={true,nocompatibility},
            final,
            tracking=true,
            kerning=true,
            spacing=true,
            factor=1100,
            stretch=10,
            shrink=10]{} %microtype

\usepackage[english]{babel}
\usepackage{blindtext}
\usepackage{helvet}
\usepackage[T1]{fontenc}
\usepackage{booktabs}
\usepackage{tabu}

\usepackage{amsmath}
\usepackage{amssymb}
\usepackage{amsthm}
\usepackage{algorithm}
\usepackage{siunitx}
\usepackage{amsmath,amsfonts,amsthm,mathrsfs, mathtools}
\usepackage{multirow}

\usepackage{marvosym}
\usepackage{graphicx}
\usepackage{pgfpages}
\graphicspath{{images/}{../images/}}
\usepackage{pgffor}
\usepackage{xcolor}
\usepackage{colortbl}
\usepackage{pgfplots}
\usepackage{tikz}
\usepackage{tikzsymbols}
\usetikzlibrary{automata,positioning,fadings,shadows, shapes.arrows, shapes.geometric}
\usetikzlibrary{matrix, chains, patterns}
\usetikzlibrary{arrows,shapes.gates.logic.US,shapes.gates.logic.IEC,calc,automata}
\usepgfplotslibrary{groupplots}

\usepackage{pifont}

\definecolor{Paired-2}{RGB}{166,206,227}
\definecolor{Paired-1}{RGB}{31,120,180}
\definecolor{Paired-4}{RGB}{178,223,138}
\definecolor{Paired-3}{RGB}{51,160,44}
\definecolor{Paired-6}{RGB}{251,154,153}
\definecolor{Paired-5}{RGB}{227,26,28}
\definecolor{Paired-8}{RGB}{253,191,111}
\definecolor{Paired-7}{RGB}{255,127,0}
\definecolor{Paired-10}{RGB}{202,178,214}
\definecolor{Paired-9}{RGB}{106,61,154}
\definecolor{Paired-12}{RGB}{255,255,153}
\definecolor{Paired-11}{RGB}{177,89,40}

\usepackage{tikz-timing}[2009/05/15]

\usepackage{scalerel} % needed package to scale the pdf-images perfectly

    \pgfdeclarelayer{background}
    \pgfdeclarelayer{foreground}
    \pgfsetlayers{background,main,foreground}
    \tikzstyle{vblock} = [draw,
                          fill=dateorange!30,
                          rotate=90,
                          anchor=north west,
                          minimum width=3.5cm,
                          minimum height=.5cm,
                         rounded corners=.05cm,
                          ]

    \tikzstyle{tile} = [draw,
                        fill=dateblue!30,
                        anchor=south west,
                        minimum width=3.5cm,
                        minimum height=.5cm,
                        rounded corners=.05cm,
                        ]

    \tikzstyle{nnlayer} = [draw,
                        fill=dateblue!30,
                        anchor=south west,
                        minimum width=1cm,
                        minimum height=.5cm,
                        rounded corners=.05cm,
                        ]

    \tikzstyle{petile} = [draw,
                        fill=dateorange!30,
                        anchor=south west,
                        minimum width=1.5cm,
                        minimum height=.5cm,
                        rounded corners=.05cm,
                        ]

    \tikzstyle{arrow} = [->,
                         >=stealth,
                         thick,
                         rounded corners=.1cm,
                         color=datemagenta,
                         ]

    \tikzstyle{arrowg} = [->,
                          >=stealth,
                          thick,
                          rounded corners=.1cm,
                          color=lightgray,
                          ]
    \tikzset{%
    table/.style={%
      matrix of nodes,
      row sep=-\pgflinewidth,
      column sep=-\pgflinewidth,
      nodes={rectangle,draw=black,text width=1.25ex,align=center},
      text depth=0.25ex,
      text height=1ex,
      nodes in empty cells
      },
    texto/.style={font=\footnotesize\sffamily},
    title/.style={font=\small\sffamily}
    }

    \pgfmathdeclarefunction{gauss}{2}{%
        \pgfmathparse{1/(#2*sqrt(2*pi))*exp(-((x-#1)^2)/(2*#2^2))}% chktex 36
    }

    \pgfmathdeclarefunction{gaussprune}{2}{%
        \pgfmathparse{or(x<-1,x>1)/(#2*sqrt(2*pi))*exp(-((x-#1)^2)/(2*#2^2))}% chktex 36
    }

\usepackage{stfloats}

\usepackage{perpage}
\usepackage{hyperref}
\usepackage{csquotes}

\usepackage{pgfpages}
\makeatother

\definecolor{dateblue}{RGB}{36,80,117}
\definecolor{datemagenta}{RGB}{183,50,101}
\definecolor{dateorange}{RGB}{198,84,54}
\definecolor{datebrown}{RGB}{198,140,54}
\definecolor{dateyellow}{RGB}{198,178,54}

\tikzfading[name=arrowfading, top color=transparent!0, bottom color=transparent!95]
\tikzset{arrowfill/.style={#1, general shadow={fill=black, shadow yshift=-0.8ex, path fading=arrowfading}}}
\tikzset{arrowstyle/.style n args={3}{draw=#2,arrowfill={#3}, single arrow,minimum height=#1, single arrow,
single arrow head extend=.3cm,}}

\NewDocumentCommand{\tikzfancyarrow}{O{2cm} O{dateblue} O{top color=dateblue!20, bottom color=dateblue} m}{
\tikz[baseline=-0.5ex]\node [arrowstyle={#1}{#2}{#3}] {#4};
}

\usepackage{circuitikz}

{
\begin{tikzpicture}[baseline=(current bounding box.north),scale=0.8]
\draw (0,0) grid (4,4);
\draw (0,4) -- node [pos=0.7,above right,anchor=south west] {$x_3x_4$} node [pos=0.7,below left,anchor=north east] {$x_1x_2$} ++(135:1);
\matrix (mapa) [matrix of nodes,
        column sep={0.8cm,between origins},
        row sep={0.8cm,between origins},
        every node/.style={minimum size=0.3mm},
        anchor=8.center,
        ampersand replacement=\&] at (0.5,0.5)
{
                       \& |(c00)| 00         \& |(c01)| 01         \& |(c11)| 11         \& |(c10)| 10         \& |(cf)| \phantom{00} \\
|(r00)| 00             \& |(0)|  \phantom{0} \& |(1)|  \phantom{0} \& |(3)|  \phantom{0} \& |(2)|  \phantom{0} \&                     \\
|(r01)| 01             \& |(4)|  \phantom{0} \& |(5)|  \phantom{0} \& |(7)|  \phantom{0} \& |(6)|  \phantom{0} \&                     \\
|(r11)| 11             \& |(12)| \phantom{0} \& |(13)| \phantom{0} \& |(15)| \phantom{0} \& |(14)| \phantom{0} \&                     \\
|(r10)| 10             \& |(8)|  \phantom{0} \& |(9)|  \phantom{0} \& |(11)| \phantom{0} \& |(10)| \phantom{0} \&                     \\
|(rf) | \phantom{00}   \&                    \&                    \&                    \&                    \&                     \\
};
}%
{
\end{tikzpicture}
}

{
\begin{tikzpicture}[baseline=(current bounding box.north),scale=0.8]
\draw (0,0) grid (4,2);
\draw (0,2) -- node [pos=0.7,above right,anchor=south west] {$x_2x_3$} node [pos=0.7,below left,anchor=north east] {$x_1$} ++(135:1);
\matrix (mapa) [matrix of nodes,
        column sep={0.8cm,between origins},
        row sep={0.8cm,between origins},
        every node/.style={minimum size=0.3mm},
        anchor=4.center,
        ampersand replacement=\&] at (0.5,0.5)
{
                      \& |(c00)| 00         \& |(c01)| 01         \& |(c11)| 11         \& |(c10)| 10         \& |(cf)| \phantom{00} \\
|(r00)| 0             \& |(0)|  \phantom{0} \& |(1)|  \phantom{0} \& |(3)|  \phantom{0} \& |(2)|  \phantom{0} \&                     \\
|(r01)| 1             \& |(4)|  \phantom{0} \& |(5)|  \phantom{0} \& |(7)|  \phantom{0} \& |(6)|  \phantom{0} \&                     \\
|(rf) | \phantom{00}  \&                    \&                    \&                    \&                    \&                     \\
};
}%
{
\end{tikzpicture}
}

{
\begin{tikzpicture}[baseline=(current bounding box.north),scale=0.8]
\draw (0,0) grid (4,2);
\draw (0,2) -- node [pos=0.7,above right,anchor=south west] {$x_1x_2$} node [pos=0.7,below left,anchor=north east] {$x_5$} ++(135:1);
\matrix (mapa) [matrix of nodes,
        column sep={0.8cm,between origins},
        row sep={0.8cm,between origins},
        every node/.style={minimum size=0.3mm},
        anchor=4.center,
        ampersand replacement=\&] at (0.5,0.5)
{
                      \& |(c00)| 00         \& |(c01)| 01         \& |(c11)| 11         \& |(c10)| 10         \& |(cf)| \phantom{00} \\
|(r00)| 0             \& |(0)|  \phantom{0} \& |(1)|  \phantom{0} \& |(3)|  \phantom{0} \& |(2)|  \phantom{0} \&                     \\
|(r01)| 1             \& |(4)|  \phantom{0} \& |(5)|  \phantom{0} \& |(7)|  \phantom{0} \& |(6)|  \phantom{0} \&                     \\
|(rf) | \phantom{00}  \&                    \&                    \&                    \&                    \&                     \\
};
}%
{
\end{tikzpicture}
}

{
\begin{tikzpicture}[baseline=(current bounding box.north),scale=0.8]
\draw (0,0) grid (4,1);
\draw (0,1) -- node [pos=0.7,above right,anchor=south west] {$x_1x_2$} node [pos=0.7,below left,anchor=north east] {} ++(135:1);
\matrix (mapa) [matrix of nodes,
        column sep={0.8cm,between origins},
        row sep={0.7cm,between origins},
        every node/.style={minimum size=0.3mm},
        anchor=5.center,
        ampersand replacement=\&] at (0.5,0.8)
{
 \& |(c00)| 00         \& |(c01)| 01         \& |(c10)| 10         \& |(c11)| 11         \& |(cf)| \phantom{00} \\
 |(rf) | \phantom{00}   \& |(0)|  \phantom{0} \& |(1)|  \phantom{0} \& |(2)|  \phantom{0} \& |(3)|  \phantom{0} \&                     \\
};
}%
{
\end{tikzpicture}
}
{
\begin{tikzpicture}[baseline=(current bounding box.north),scale=0.8]
\draw (0,0) grid (4,1);
\draw (0,1) -- node [pos=0.7,above right,anchor=south west] {$x_1x_2$} node [pos=0.7,below left,anchor=north east] {} ++(135:1);
\matrix (mapa) [matrix of nodes,
        column sep={0.8cm,between origins},
        row sep={0.7cm,between origins},
        every node/.style={minimum size=0.3mm},
        anchor=5.center,
        ampersand replacement=\&] at (0.5,0.8)
{
 \& |(c00)| 00         \& |(c01)| 01         \& |(c11)| 11         \& |(c10)| 10         \& |(cf)| \phantom{00} \\
 |(rf) | \phantom{00}   \& |(0)|  \phantom{0} \& |(1)|  \phantom{0} \& |(3)|  \phantom{0} \& |(2)|  \phantom{0} \&                     \\
};
}%
{
\end{tikzpicture}
}

{
\begin{tikzpicture}[baseline=(current bounding box.north),scale=0.8]
\draw (0,0) grid (2,2);
\draw (0,2) -- node [pos=0.7,above right,anchor=south west] {$x_2$} node [pos=0.7,below left,anchor=north east] {$x_1$} ++(135:1);
\matrix (mapa) [matrix of nodes,
        column sep={0.8cm,between origins},
        row sep={0.8cm,between origins},
        every node/.style={minimum size=0.3mm},
        anchor=2.center,
        ampersand replacement=\&] at (0.5,0.5)
{
          \& |(c00)| 0          \& |(c01)| 1  \\
|(r00)| 0 \& |(0)|  \phantom{0} \& |(1)|  \phantom{0} \\
|(r01)| 1 \& |(2)|  \phantom{0} \& |(3)|  \phantom{0} \\
};
}%
{
\end{tikzpicture}
}

\usepackage{listings}
\usetikzlibrary{positioning, shapes, arrows.meta}

\usepackage[utf8x]{inputenc}

\definecolor{shellGreen}{RGB}{19,193,106}
\definecolor{backcolor}{rgb}{0.95,0.95,0.92}
\definecolor{mateBlack}{RGB}{45,45,50}
\definecolor{comment}{rgb}{0.1,0.6,0.2}
\definecolor{codegray}{rgb}{0.5,0.5,0.5}

\lstdefinestyle{verilog}{
   language=verilog,
   frame=single,
   basicstyle=\ttfamily\footnotesize,
   breaklines=true,
   captionpos=t,
   keepspaces=true,
   backgroundcolor=\color{backcolor},
   keywordstyle=[1]\color{blue}\bf,
   keywordstyle=[2]\color{red}\bf,
   keywordstyle=[3]\color{cyan!50}\bf,
   stringstyle=\color{orange},
   commentstyle=\color{comment},
   tabsize=2,
   showspaces=false,
   showstringspaces=false,
   showtabs=false,
   morekeywords=[1]{
      library, use ,all,entity,is,port,in,out,end,architecture,of, body,
      function, variable, begin,and,or,Not,downto,ALL, signal, process, if,
      else, elsif, case, when, then, range, to, component, type, with, select,
      others, constant, inout, buffer, map, true, false, array, subtype, wait,
      wait for, generic, =, <, >, <=, >=, =>, 
   },
   alsoletter={=, <, >},
   morekeywords=[2]{
          STD, textio, std_logic_textio, STD_LOGIC_VECTOR,STD_LOGIC,IEEE,STD_LOGIC_1164, work, local, real,
          math_real, time, NUMERIC_STD,STD_LOGIC_ARITH,STD_LOGIC_UNSIGNED,
          std_logic_vector, std_logic, ieee, numeric_std, std_ulogic,
          std_logic_1164, natural, bit, bit_vector, signed, unsigned,
          boolean, integer
    },
    morekeywords=[3]{rising_edge, falling_edge, resize, to_signed, to_unsigned},
    morecomment=[l]{--},
    rulecolor=\color{black},
}

\usepackage{diagbox}

\definecolor{codegreen}{rgb}{0,0.6,0}
\definecolor{codegray}{rgb}{0.5,0.5,0.5}
\definecolor{codepurple}{rgb}{0.58,0,0.82}
\definecolor{backcolour}{rgb}{0.95,0.95,0.92}

\lstdefinestyle{mystyle}{
    backgroundcolor=\color{backcolour}\bf,   
    commentstyle=\color{codegreen}\bf,
    keywordstyle=\color{magenta}\bf,
    numberstyle=\tiny\color{codegray}\bf,
    stringstyle=\color{codepurple},
    basicstyle=\ttfamily\footnotesize,
    breakatwhitespace=false,         
    breaklines=true,                 
    captionpos=t,                    
    keepspaces=true,                
    showspaces=false,                
    showstringspaces=false,
    showtabs=false,                  
    tabsize=2
}

%% file: 00_abstract.tex
\begin{abstract}
  Efficient sampling of satisfying formulas for circuit satisfiability (CircuitSAT), a well-known NP-complete problem, is essential in modern front-end applications for thorough testing and verification of digital circuits. Generating such samples is a hard computational problem due to the inherent complexity of digital circuits, size of the search space, and resource constraints involved in the process. Addressing these challenges has prompted the development of specialized algorithms that heavily rely on heuristics. However, these heuristic-based approaches frequently encounter scalability issues when tasked with sampling from a larger number of solutions, primarily due to their sequential nature. Different from such heuristic algorithms, we propose a novel differentiable sampler for multi-level digital circuits, called {\sc Demotic}, that utilizes gradient descent (GD) to solve the CircuitSAT problem and obtain a wide range of valid and distinct solutions. {\sc Demotic} leverages the circuit structure of the problem instance to learn valid solutions using GD by re-framing the CircuitSAT problem as a supervised multi-output regression task. This differentiable approach allows bit-wise operations to be performed independently on each element of a tensor, enabling parallel execution of learning operations, and accordingly, GPU-accelerated sampling with significant runtime improvements compared to state-of-the-art heuristic samplers. We demonstrate the superior runtime performance of {\sc Demotic} in the sampling task across various CircuitSAT instances from the ISCAS-85 benchmark suite. Specifically, {\sc Demotic} outperforms the state-of-the-art sampler by more than two orders of magnitude in most cases.
\end{abstract}

% \keywords{Circuit Satisfiability, Gradient Descent, Multi-level Circuits, Verification, and Testing.}

\begin{IEEEkeywords}
Circuit Satisfiability, Gradient Descent, Multi-level Circuits, Verification, and Testing.
\end{IEEEkeywords}

%% file: 01_introduction.tex
\section{Introduction}
Circuit satisfiability (CircuitSAT) solving is an integral part of testing and verification of digital circuits in modern front-end applications such as logic rewriting, false path analysis, property checking, logic synthesis and equivalence checking \cite{Mishchenko2005Optimization, Tsai2009TimingAnalyzer, Bradley2011ModelChecking, Mishchenko2006EquivalenceChecking, Zhang2021LogicSynthesis}. CircuitSAT samplers play a crucial role in generating diverse samples from the solution space, aiding in validation, analysis, and optimization of digital circuit designs \cite{dutra2018quicksampler}. Exhaustive exploration and the creation of diverse solutions are essential to guaranteeing that the design fulfills its functional requirements and operates correctly across different scenarios. Consequently, numerous sampling techniques have been developed to detect edge cases and outliers, ensure representativeness, bolster robustness, and promote the broad applicability of findings \cite{dutra2019EfficientSampling}.

High-throughput sampling is fundamental in the realm of CircuitSAT, playing a vital role in various critical tasks such as constrained random verification (CRV) \cite{Kitchen2007crv}. Its primary function lies in boosting efficiency and scalability by enabling swift exploration of vast solution spaces, which is especially crucial when dealing with complex digital circuits containing numerous inputs, outputs, and intermediate signals. Moreover, it expands coverage across solution spaces, assisting in the detection of uncommon solutions and intricate edge cases inherent in CircuitSAT problems. Furthermore, high-throughput sampling enhances statistical reliability by producing larger sample sizes, thereby reducing sampling variability in CircuitSAT formula analysis.

A common approach for solving the CircuitSAT problem and obtaining diverse solutions involves transforming the CircuitSAT problem into a Boolean Satisfiability (SAT) problem, then employing robust and advanced solvers to solve it effectively \cite{Hsu2014CircuitSAT}. More precisely, the CircuitSAT solving process begins with the formulation of the logical constraints of a digital logic into a Boolean formula, typically represented in conjunctive normal form (CNF). This formula encodes the functionality of the underlying circuits into a set of Boolean clauses according to the interconnections between the circuit's inputs, outputs and internal signals, ensuring that the resulting CNF accurately represents the original circuit's behavior \cite{Velev2004CNF}. 

The conversion process can sometimes be complex and computationally intensive, especially for large circuits. In some cases, the resulting CNF might not always be \textit{compact} due to different factors such as the size of the circuit and the number of intermediate variables. For instance, circuits containing many gates or components may result in a large CNF since each gate or component can potentially introduce additional variables and clauses. The intermediate variables can also introduce additional variables and clauses, contributing in the size of the resulting CNF. This increase in complexity of CNF can impact the efficiency of SAT solvers. 

SAT solvers employs a variety of techniques to search for a satisfying assignment to the variables in the CNF. Modern SAT solvers \cite{Niklas2003SAT, Moskewicz2001Chaff, Audemard2018Glucose} often utilize conflict-driven clause learning (CDCL) algorithm \cite{Silva1996CDCL, silva2021CDCL} which heavily relies on heuristics such as conflict-driven backtracking and clause learning. Theses heuristics help guide the search process of CDCL for a satisfying assignment efficiently. Due to the sequential nature of these heuristics and their reliance on branching and backtracking, current state-of-the-art (SOTA) SAT solvers are executed on CPUs. Accordingly, SOTA SAT samplers, which integrate a SAT solver as part of their algorithms, depend on a sequential process and are also optimized for execution on CPUs.

GPU acceleration has demonstrated substantial throughput performance advantages across diverse applications especially in machine learning \cite{Krizhevsky2012AlexNet}. In general, algorithms that exhibit parallelism and can be decomposed into independent tasks are generally suitable for GPU acceleration. This appears to align with the requirements for generating various satisfying solutions to the CircuitSAT problem, if there was a sampling method performing regular and data-parallel computations. In addition to parallelism, introducing a sampling method that operates directly on circuits by leveraging the spatial and temporal dimensions of digital computations—without converting to CNF—is crucial, as the CNF conversion process may not consistently produce the most optimal or compact representation. To this end, we introduce a novel differentiable sampler for multi-level digital circuits, called {\sc Demotic} \footnote{The code of {\sc Demotic} is available at \url{https://github.com/arashardakani/Demotic}.}, that utilizes gradient descent (GD) for learning diverse solutions to the CircuitSAT problem. We re-frame the CircuitSAT problem as a multi-output regression task, where each logic gate is modeled with a probabilistic representation. We subsequently formulate a loss function by incorporating specified constraints into the circuit. This approach allows us to maintain the integrity of the circuit structure while transforming the sampling process into a learning process. This process enables the generation of independent solutions to the CircuitSAT problem in a parallel fashion, enabling acceleration with GPUs. In summary, we make the following contributions in this paper.

\begin{itemize}
    \item We introduce a novel differentiable sampler for multi-level digital circuits, called {\sc Demotic}.
    \item We use a probabilistic representation to model logic gates and convert the CircuitSAT problem into a multi-output regression task.
    \item Our proposed sampling method maintains the original structure of the underlying digital circuit without the need for the conversion into any Boolean formulation.
    \item {\sc Demotic} enables performing the learning process in parallel, leading to GPU-accelerated sampling of satisfying formulas for the CircuitSAT problem.
    \item We demonstrate the performance of {\sc Demotic} across different CircuitSAT instances from the ISCAS-85 benchmark suite \cite{Hansen1999ISCASBench}. 
    
\end{itemize}

%% file: 04_preliminaries.tex
\vspace{-0.25cm}
\section{Preliminaries}
\subsection{CircuitSAT Sampling} \label{subsec:circuitsat_sampling}
CircuitSAT sampling refers to the task of sampling solutions from the solution space of a given CircuitSAT problem. In a Boolean circuit, variables can only take on binary values of either $0$ or $1$. A digital circuit is composed of various logic gates such as AND, OR, and NOT gates, which manipulate these Boolean variables. The output of such a circuit is produced based on how these logic gates operate.

In CircuitSAT, the objective is to determine whether a given circuit, representing a Boolean expression, has an assignment of binary values to its variables that results in the circuit output valuation to $1$. The sampling aspect introduces a probabilistic dimension to this problem. Instead of finding a single solution for satisfiable problems, CircuitSAT sampling aims to generate multiple solutions or samples from the set of all possible solutions. Sampling solutions from CircuitSAT instances is an integral part of the design verification process, with significant applications in CRV \cite{Kitchen2007crv}.

CRV is a verification methodology employed in the design and testing of modern digital circuits. It involves generating random input stimuli for the design under test (DUT) while adhering to a set of predefined constraints. This method helps explore a wide range of possible input scenarios, increasing the likelihood of identifying design flaws and ensuring robustness. Inputs to the DUT are generated randomly within specified constraints. This randomness helps cover a broad spectrum of test scenarios, including corner cases that might not be easily detected with directed tests. Constraints are rules or conditions that the random inputs must satisfy. These can be functional constraints based on the design specifications or operational constraints based on practical considerations. Constraints ensure that the generated random inputs are valid and meaningful for the DUT.

For instance, consider a 4-bit multiplier with two unsigned 4-bit binary inputs and an 8-bit product as its primary output. To restrict our bug search to input pairs where the product is less than $128$, we need to set the most significant bit of the product to $0$ and find random satisfying solutions. It is worth mentioning that while exhaustive search can be used for design verification with complete coverage, it is often impractical for complex designs due to scalability and efficiency limitations. CRV provides a more feasible, efficient, and effective approach by leveraging constraints and randomization to achieve high coverage and uncover critical issues within a manageable timeframe.

One common approach for CircuitSAT sampling is the use of SAT solvers with sampling capabilities. These solvers are designed to not only determine the satisfiability of a Boolean formula but also to sample solutions from the solution space. There are various technique for efficient SAT solving, including backtracking algorithms like Davis-Putnam-Logemann-Loveland (DPLL) algorithm \cite{Davis1962DPLL}, stochastic local search methods like WalkSAT \cite{selman1993local}, and CDCL algorithms \cite{Silva1996CDCL, silva2021CDCL}. Over the past years, various algorithms and techniques have been developed for CircuitSAT/SAT sampling including randomized algorithms, Markov chain Monte Carlo (MCMC) methods, and sampling-based heuristics \cite{Impagliazzo2017RandomSAT, kitchen2009markov, Soos2020unigen3, dutra2018quicksampler, Golia2021cmsgen}. These approaches typically involve iteratively exploring the solution space, selecting candidate solutions based on certain criteria, and stochastically accepting or rejecting them.

The process of CircuitSAT sampling using SAT solvers involves translating the structure and logic of the given circuit into an equivalent Boolean formula, typically represented in CNF. The CNF consists of a conjunction of multiple clauses (i.e., AND of multiple clauses), where each clause is a disjunction of literals (i.e., OR of literals). Literals are referred to as Boolean variables or their complements. During the conversion process, the number of inputs, outputs and intermediate signals directly contributes to the number of variables in the CNF. The functionality of logic gates and their interconnections determine the number of clauses in the CNF. 

The size and complexity of the CNF formula vary significantly depending on factors such as the number of gates, the depth of the circuit, and the number of inputs, outputs and intermediate signals. In general, the CNF formula tends to contain significantly more bit-wise operations than its corresponding circuit. The introduced complexity to the SAT instance as the result of the conversion exponentially increases the time required to find a solution using SAT solvers since the SAT problem is NP-complete. This issue becomes worse when the complexity of digital circuits increases, making SAT sampling a non-trivial task, especially for large or complex circuits.

\vspace{-0.25cm}
\subsection{Multi-Output Regression Task}
A multi-output regression task is a statistical technique used to predict multiple target variables simultaneously from a set of input variables \cite{borchani2015survey}. In this task, the primary objective is to develop a model that accurately captures the relationships between input and output variables. The model can be constructed using various techniques such as linear regression and neural networks. The model is then trained using a dataset where both input-output pairs are known. During training, the model's parameters are adjusted by minimizing the distance between the predicted outputs and the desired target variables. A common metric to measure such a distance is mean squared error (MSE) or $\ell_2$-loss.

%% file: 03_methodology.tex
\section{Methodology}

\begin{figure*}[t]
    \centering
    \scalebox{0.9}{
    \input{figure1}}
    \caption{An overview of {\sc Demotic} is shown. {\sc Demotic} takes a Verilog instance describing a combinational circuit and parse it into its corresponding probabilistic model described in PyTorch. The embedding layer converts the learnable real-value inputs into probabilities. The $\ell_2$-loss function is calculated in each training iteration and the input variables are updated using GD.}
    \label{fig1}
\end{figure*}
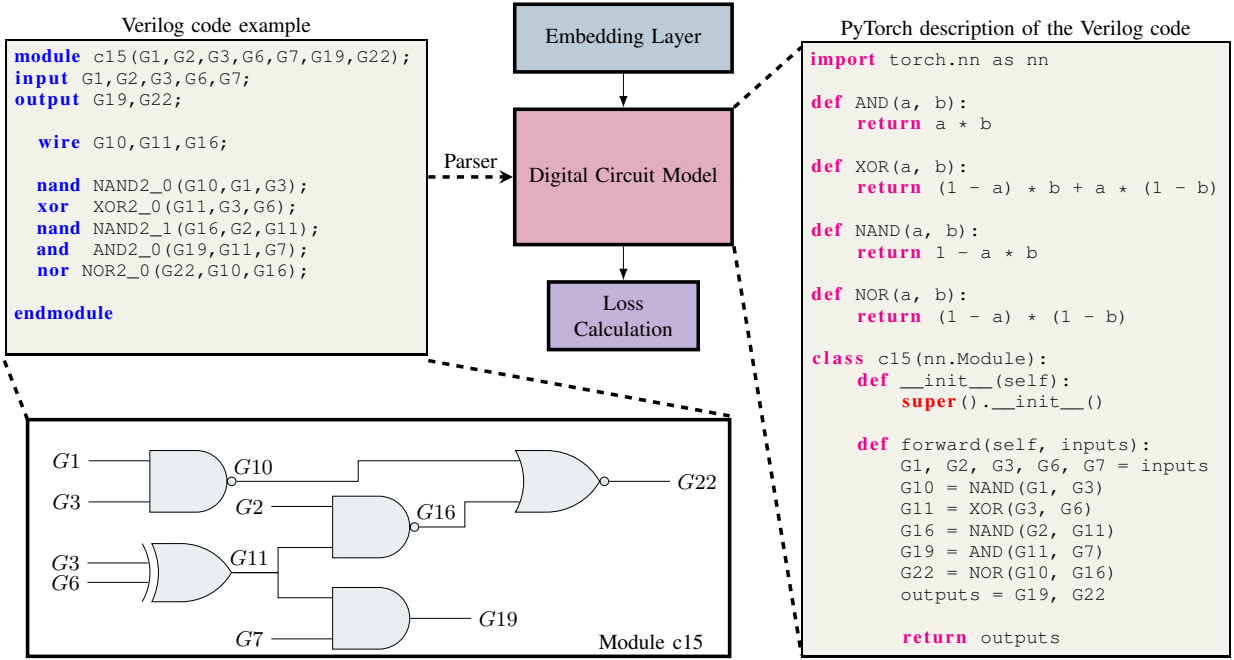

In this section, we describe our differentiable solver/sampler for multi-level digital circuits. While the common approach in solving CircuitSAT typically involves converting the underlying circuit into CNF and employing a SAT solver to find the satisfying solution, we take a completely different approach. Instead, we re-frame the CircuitSAT problem as a multi-output regression task, transforming it into a learning problem. Digital circuits are inherently discrete and non-differentiable. Therefore, we first need to relax the CircuitSAT problem into a continuous form while accurately capturing the structure and behavior of the circuit. To accomplish this, we leverage the probability model of digital gates, as shown in Table \ref{tab1}. This probability model is commonly used in different domains such as stochastic computing \cite{Ardakani2017SC} and dynamic power estimation of digital circuits \cite{harris2010cmos}. We use these probabilities to model each gate in the circuit. The result of such modeling is a differentiable formulation of the underlying circuit that accurately describes its functionality while preserving its spatial structure. Of course, the outcome of this model remains identical to the original circuit in its discrete form for any binary input valuations.

\begin{table}[t]
    \centering
    \caption{Probability model of logic gates.}
    \vspace{-0.25cm}
    \input{table1}
    \label{tab1}
    \vspace{-0.25cm}
\end{table}

Given the differentiable model of the circuit obtained by replacing its discrete logic gates with their corresponding probability model, our objective now is to generate a set of inputs that satisfy a desired constraint. This constraint could pertain to any desired valuation of intermediate signals or outputs. To generate satisfying solutions to the CircuitSAT problem, we represent the input variables to the circuit as $\textbf{V} \in \mathbb{R}^{b\times n}$, where $n$ represents the number of variables and $b$ denotes the batch size. We define the matrix $\textbf{V}$ as the parameters of an embedding layer in our circuit model, which will be updated during the learning process. It is worth mentioning that the number of variables in our sampling method is significantly fewer than that of SAT samplers, remaining the same as the number of inputs in the circuit. This discrepancy arises because SAT samplers deal with the CNF of the circuit, where each gate or component introduces additional variables. The embedding layer converts the real-value input variables of the circuit into probabilities in the range from $0$ to $1$ using the sigmoid function $\sigma(\cdot)$, expressed as:
\begin{equation}
    \textbf{P} = \sigma(\textbf{V}) = \dfrac{1}{1 + e^{-\textbf{V}}},
\end{equation}
where $\textbf{P} \in [0, 1]^{b\times n}$ represents the input probabilities to the underlying circuit. The circuit functionality is then computed as:
\begin{equation}
    \textbf{Y} = \mathcal{F}(\textbf{P}),
\end{equation}
where $\mathcal{F}:[0, 1]^{b \times n} \rightarrow [0, 1]^{b \times m}$ denotes the probabilistic model of the circuit. The matrix $\textbf{Y} \in [0, 1]^{b \times m}$ denotes the $m$ outputs across $b$ data batches. The $\ell_2$-loss function $\mathcal{L}$ can be constructed by measuring the distance between $\textbf{Y}$ and the target output valuation matrix $\textbf{T} \in \{0, 1\}^{b \times m}$ as follows:
\begin{equation}
    \mathcal{L} = \sum_{b,m} \left|\left| \textbf{Y} - \textbf{T} \right|\right|^2_2.
\end{equation}
The above loss function can be minimized, and the input variables (i.e., $\textbf{V}$) can be updated using GD in an iterative manner. Upon convergence, the $b$ solutions to the CircuitSAT problem are obtained by converting the soft input values (i.e., $\textbf{V}$) into hard values (i.e., $\widetilde{\textbf{V}} \in \{0, 1\}^{b\times n}$).

Fig. \ref{fig1} illustrates the overview of {\sc Demotic}. {\sc Demotic} is equipped with a parser to covert the circuit described in either bit-blasted Verilog or Berkeley Logic Interchange Format (BLIF) into its corresponding probabilistic model. Consequently, {\sc Demotic} can describe combinational circuits and generate satisfying solutions for any arbitrary constraint on the circuit. Such a sampling paradigm can also benefit from GPU acceleration due to the parallel independent computations across the data batches, enabling a high-throughput sampling procedure. 

To better understand our methodology, let us consider a quantitative example using the module ``c$15$'' shown in Fig. \ref{fig1}. We set the output node $G19$ to $1$ as an output constraint, while the output node $G22$ can take any value of either $0$ or $1$. Therefore, the goal in this example is to find a pair of inputs such that the output node $G19$ is equal to $1$. In this example, the input nodes contributing to our output constraint are $G3$, $G6$, and $G7$. These inputs are learned iteratively using gradient descent. The remaining input nodes, $G1$, $G2$, and $G3$, will not be updated and can take any arbitrary binary values. During each training iteration, each input node is updated by computing the derivative of the loss function with respect to each input node.

To illustrate the process, we generate two samples. In the first step, we randomly assign two values to each input node as follows:
\begin{equation}
    \textbf{v}_{G3} = \begin{bmatrix}
           0.1 \\
           -0.2 
         \end{bmatrix}, \textbf{v}_{G6} = \begin{bmatrix}
           0.5 \\
           -0.4 
         \end{bmatrix}, \textbf{v}_{G7} = \begin{bmatrix}
           -0.7 \\
           -0.8 
         \end{bmatrix},
\end{equation}
where the concatenation of the above vectors forms the matrix $\textbf{V}$. Next, the input probabilities to the circuit are calculated using the sigmoid function:
\begin{equation}
    \textbf{p}_{G3} = \begin{bmatrix}
           0.5250 \\
           0.4502
         \end{bmatrix}, \textbf{p}_{G6} = \begin{bmatrix}
           0.6225 \\
           0.4013
         \end{bmatrix}, \textbf{p}_{G7} = \begin{bmatrix}
           0.3318 \\
           0.3100
         \end{bmatrix}.
\end{equation}
Using the probability model of each gate shown in Table \ref{tab1}, the probabilities of the intermediate node $G11$ and the output node $G19$ are calculated as follows:
\begin{equation}
    \textbf{p}_{G11} = \begin{bmatrix}
           0.4939 \\
           0.4902
         \end{bmatrix}, \textbf{p}_{G19} = \begin{bmatrix}
           0.1639 \\
           0.1520
         \end{bmatrix}.
\end{equation}
Given the target value of 1 for the output node $G19$, the loss is calculated as:
\begin{equation}
    \mathcal{L} = (\textbf{p}_{G19} - 1)^2 = \begin{bmatrix}
           (0.1639 - 1)^2  \\
           (0.1520 - 1)^2 
         \end{bmatrix} = \begin{bmatrix}
           0.6991  \\
           0.7192 
         \end{bmatrix}.
\end{equation}

The above computations are commonly referred to as forward computations. To update the value of the input variables, we need to calculate the derivative of the loss with respect to each input variable, which is referred to as backward computations. This involves using the derivatives of each gate (as shown in Table \ref{tab1}) and applying the chain rule. The process is derived as follows:
\begin{align}
    \dfrac{\partial \mathcal{L}}{\partial \textbf{v}_{G3}} &= \dfrac{\partial \mathcal{L}}{\partial \textbf{p}_{G19}} \dfrac{\partial \textbf{p}_{G19}}{\partial \textbf{p}_{G11}} \dfrac{\partial \textbf{p}_{G11}} {\partial \textbf{p}_{G3}}
    \dfrac{\partial \textbf{p}_{G3}} {\partial \textbf{v}_{G3}} = 2\textbf{p}_{G19} \cdot \textbf{p}_{G7} \cdot (1 - 2\textbf{p}_{G6}) \nonumber \\ 
    &\cdot \sigma(\textbf{v}_{G3})\cdot (1 - \sigma(\textbf{v}_{G3})) = \begin{bmatrix}
           0.0339  \\
           -0.0257 
         \end{bmatrix}, \nonumber 
\end{align}
\begin{align}
    \dfrac{\partial \mathcal{L}}{\partial \textbf{v}_{G6}} &= \dfrac{\partial \mathcal{L}}{\partial \textbf{p}_{G19}} \dfrac{\partial \textbf{p}_{G19}}{\partial \textbf{p}_{G11}} \dfrac{\partial \textbf{p}_{G11}} {\partial \textbf{p}_{G6}}
    \dfrac{\partial \textbf{p}_{G6}} {\partial \textbf{v}_{G6}} = 2\textbf{p}_{G19} \cdot \textbf{p}_{G7} \cdot (1 - 2\textbf{p}_{G3})  \nonumber 
 \\ 
    & \cdot \sigma(\textbf{v}_{G6}) \cdot (1 - \sigma(\textbf{v}_{G6}))  = \begin{bmatrix}
           0.0065   \\
           -0.0126 
         \end{bmatrix}, \nonumber 
\end{align}
\begin{align}
    \dfrac{\partial \mathcal{L}}{\partial \textbf{v}_{G7}} &= \dfrac{\partial \mathcal{L}}{\partial \textbf{p}_{G19}} \dfrac{\partial \textbf{p}_{G19}}{\partial \textbf{p}_{G7}} \dfrac{\partial \textbf{p}_{G7}} {\partial \textbf{v}_{G7}} = 2\textbf{p}_{G19} \cdot \textbf{p}_{G11} \nonumber 
 \\ 
    & \cdot \sigma(\textbf{v}_{G7})\cdot (1 - \sigma(\textbf{v}_{G7})) = \begin{bmatrix}
           -0.1831   \\
           -0.1778 
         \end{bmatrix},
\end{align}
where ``$\cdot$'' denotes element-wise multiplication.

At this point, each variable is updated using the gradient descent update rule. This involves subtracting the derivative of the loss, scaled by the learning rate, from the corresponding input variables. Given a learning rate of $\gamma = 10$, the new values of the input variables at the end of this iteration are obtained as follows:
\begin{align}
    \textbf{v}_{G3} &= \textbf{v}_{G3} - \gamma \dfrac{\partial \mathcal{L}}{\partial \textbf{v}_{G3}} =  \begin{bmatrix}
           -0.2389 \\
           0.0569
         \end{bmatrix}, \textbf{v}_{G6} = \begin{bmatrix}
           0.4349 \\
           -0.2741
         \end{bmatrix}, \nonumber \\ \textbf{v}_{G7} & = \begin{bmatrix}
           1.1311 \\
           0.9783
         \end{bmatrix}.
\end{align}
This process can be repeated multiple times until convergence. However, even after one iteration in this specific example, we obtain two valid and distinct solutions by rounding the input variables to their nearest discrete values after applying the sigmoid function. In this example, the two input pairs of $(v_{G3} = -0.2389, v_{G6} = 0.4349, v_{G7} = 1.1311)$ and $(v_{G3} = 0.0569, v_{G6} = -0.2741, v_{G7} = 0.9783)$ are rounded to $(\widetilde{v}_{G3} = 0, \widetilde{v}_{G6} = 1, \widetilde{v}_{G7} = 1)$ and $(\widetilde{v}_{G3} = 1, \widetilde{v}_{G6} = 0, \widetilde{v}_{G7} = 1)$, respectively. As demonstrated through this example, the forward and backward computations of the two samples are independent of each other. This allows for the parallel execution of the learning process across multiple samples (i.e., batches), enabling GPU acceleration.

%% file: figure1.tex
\begin{tikzpicture}[node distance=2cm,>=Latex]
    \node [ text width=6cm, align=center, line width=1.5pt] (code) {
        \begin{lstlisting}[title = Verilog code example]
module c15(G1,G2,G3,G6,G7,G19,G22);
input G1,G2,G3,G6,G7;
output G19,G22;

  wire G10,G11,G16;

  nand NAND2_0(G10,G1,G3);
  xor  XOR2_0(G11,G3,G6);
  nand NAND2_1(G16,G2,G11);
  and  AND2_0(G19,G11,G7);
  nor NOR2_0(G22,G10,G16);

endmodule
        \end{lstlisting}
    };
    
    \node [draw,  fill = datemagenta!40, text width=3cm, minimum height=2cm, align=center, right=1.25cm of code, line width=1.5pt] (model) {Digital Circuit Model};
    \node [draw,  text width=3cm,fill = dateblue!30, minimum height=1cm, align=center, above=0.5cm of model, line width=1.5pt] (embedding) {Embedding Layer};
    \node [draw, fill = Paired-9!40, text width=2cm, minimum height=1cm,align=center, below=0.5cm of model, line width=1.5pt] (loss) {Loss Calculation};
    \node [text width=6.1cm, align=center, right=1.cm of model, yshift=-2.25cm] (parsed) { \lstset{style=mystyle} \begin{lstlisting}[language=Python, title=PyTorch description of the Verilog code]
import torch.nn as nn

def AND(a, b):
    return a * b

def XOR(a, b):
    return (1 - a) * b + a * (1 - b)

def NAND(a, b):
    return 1 - a * b

def NOR(a, b):
    return (1 - a) * (1 - b)
    
class c15(nn.Module):
    def __init__(self):
        super().__init__()

    def forward(self, inputs):
        G1, G2, G3, G6, G7 = inputs
        G10 = NAND(G1, G3)
        G11 = XOR(G3, G6)
        G16 = NAND(G2, G11)
        G19 = AND(G11, G7)
        G22 = NOR(G10, G16)
        outputs = G19, G22

        return outputs
\end{lstlisting}};
    
    % Arrows
    \draw [->, >=stealth, line width=1.5pt, dashed] (code) -- node[above] {Parser} (model);
    \draw [->] (embedding.south) -- (model.north);
    \draw [->] (model.south) -- (loss.north);
    \draw [dashed, >=stealth, line width=1.5pt] (model.north east) -- ([shift={(0,-1cm)}]parsed.north west);
    \draw [dashed, >=stealth, line width=1.5pt] (model.south east) -- ([shift={(0,0.35cm)}]parsed.south west);

    \begin{scope}[scale=0.9] 
    
    \node[nand gate US, draw, logic gate inputs={nnnnn}, fill =  dateblue!10, scale = 0.9] (NAND2_0) at (0 -0.5,-5) {};
    \node[xor gate US, draw, logic gate inputs={nnnnn}, fill =  dateblue!10, scale = 1.7] (NAND2_1) at (0 -0.5,-6.5) {};
    \node[nand gate US, draw, logic gate inputs={nnnnn}, fill =  dateblue!10, scale = 0.9] (NAND2_2) at (3 -0.5,-5.75) {};
    \node[and gate US, draw, logic gate inputs={nnnnn}, fill =  dateblue!10, scale = 0.9] (NAND2_3) at (3 -0.5,-7.25) {};
    \node[nor gate US, draw, logic gate inputs={nnnnn}, fill =  dateblue!10, scale = 0.9] (NAND2_4) at (6 -0.5,-5) {};
    \node[] (NAND2_5) at (6 -0.5,-6.5) {};

    \draw (NAND2_0.input 1) -- ++(-1,0) node[left] {$G1$};
    \draw (NAND2_0.input 5) -- ++(-1,0) node[left] {$G3$};
    \draw (NAND2_1.input 1) -- ++(-1,0) node[left] {$G3$};
    \draw (NAND2_1.input 2) -- ++(-1,0) node[left] {$G6$};
    \draw (NAND2_2.input 1) -- ++(-1,0) node[left] {$G2$};
    % \draw (NAND2_2.input 5) -- ++(-1,0) node[left] {$G9$};
    % \draw (NAND2_3.input 1) -- ++(-1,0) node[left] {$G9$};
    \draw (NAND2_3.input 5) -- ++(-1,0) node[left] {$G7$};
    
    % \draw (NAND2_4.input 1) -- ++(-1,0) node[left] {$G8$};
    % \draw (NAND2_4.input 5) -- ++(-1,0) node[left] {$G12$};
    % \draw (NAND2_5.input 1) -- ++(-1,0) node[left] {$G12$};
    % \draw (NAND2_5.input 5) -- ++(-1,0) node[left] {$G15$};
    
    % \draw (NAND2_0.output) -- (NAND2_1.input 1);
    \draw (NAND2_1.output)  node[shift={(-0.15,0)}, above right] {$G11$} -- ([xshift=0.75cm]NAND2_1.output) |- (NAND2_2.input 5);
    \draw (NAND2_1.output) -- ([xshift=0.75cm]NAND2_1.output) |- (NAND2_3.input 1);
    \draw (NAND2_0.output) node[shift={(-0.15,0)}, above right] {$G10$}  -- ([xshift=2cm]NAND2_0.output) |- (NAND2_4.input 1);
    \draw (NAND2_2.output) node[shift={(-0.15,0)}, above right] {$G16$}  -- ([xshift=0.75cm]NAND2_2.output) |- (NAND2_4.input 5);
    % \draw (NAND2_2.output) -- (NAND2_4.input 5);
    % \draw (NAND2_3.output) -- (NAND2_5.input 5);
    % \draw (NAND2_3.output) -- ([xshift=0.75cm]NAND2_3.output) |- (NAND2_5.input 5);
    % \draw (NAND2_2.output) -- ([xshift=0.75cm]NAND2_2.output) |- (NAND2_5.input 1);
    \draw (NAND2_4.output) -- ++(1,0) node[right] {$G22$};
    \draw (NAND2_3.output) -- ++(1,0) node[right] {$G19$};

        \draw [line width=1.5pt]([shift={(-1.5,0.)}]NAND2_0.north west) ++(-0.5,0.5) rectangle ([shift={(2.8,-1.25)}]NAND2_5.south east);
        \node[above left] at ([shift={(2.5,-1.25)}]NAND2_5.south east) {Module c15};

    \draw [dashed, >=stealth, line width=1.5pt] ([shift={(0,0.29)}]code.south west) -- ([shift={(-2,0.5)}]NAND2_0.north west);
    \draw [dashed, >=stealth, line width=1.5pt] ([shift={(0,0.29)}]code.south east) -- ([shift={(2.8,2.7)}]NAND2_5.south east);
    \end{scope}
    
\end{tikzpicture}

%% file: table1.tex
\scalebox{0.77}{\begin{tabular}{c|c|c|c} 
\hline
\midrule
Gate & Input Probability & Output Probability & Derivative w.r.t Input\\ 
 \midrule
NOT & $P_{1}$ & $P_y = \overline{P_1} = 1 - P_{1}$ & $\dfrac{\partial P_y}{\partial P_1} = -1$  \\
\midrule
AND & $P_{1}$, $P_{2}$ & $P_y = P_{1}~P_{2}$ & $\dfrac{\partial P_y}{\partial P_1} = P_2$, $\dfrac{\partial P_y}{\partial P_2} = P_1$ \\
\midrule
OR & $P_{1}$, $P_{2}$ & $P_y = 1 - \overline{P_{1}}~\overline{P_{2}}$ & $\dfrac{\partial P_y}{\partial P_1} = \overline{P_{2}}$, $\dfrac{\partial P_y}{\partial P_2} = \overline{P_{1}}$\\
\midrule
NAND & $P_{1}$, $P_{2}$ & $P_y = 1 - P_{1}~P_{2}$ & $\dfrac{\partial P_y}{\partial P_1} = -P_2$, $\dfrac{\partial P_y}{\partial P_2} = -P_1$\\
\midrule
NOR & $P_{1}$, $P_{2}$ & $P_y = \overline{P_{1}}~\overline{P_{2}}$ & $\dfrac{\partial P_y}{\partial P_1} = -\overline{P_{2}}$, $\dfrac{\partial P_y}{\partial P_2} = -\overline{P_{1}}$\\
\midrule
XOR & $P_{1}$, $P_{2}$ & $P_y = \overline{P_{1}}~P_{2} + P_{1}~\overline{P_{2}}$ & $\dfrac{\partial P_y}{\partial P_1} = 1 -2P_2$, $\dfrac{\partial P_y}{\partial P_2} = 1 - 2 P_1$\\
\midrule
XNOR & $P_{1}$, $P_{2}$ & $P_y = P_{1}~P_{2} + \overline{P_{1}}~\overline{P_{2}}$ & $\dfrac{\partial P_y}{\partial P_1} = 2P_2-1$, $\dfrac{\partial P_y}{\partial P_2} = 2P_1-1$\\
\midrule
\hline

\end{tabular}}

%% file: 05_experiments.tex
\section{Experimental Results}
In this section, we showcase how our differentiable method tackles sampling in CircuitSAT problems. To achieve this, we have created a prototype for {\sc Demotic} using PyTorch. PyTorch is an open-source machine learning framework that blends Torch's efficient GPU-accelerated backend libraries with a user-friendly Python interface. For a comprehensive evaluation, we use the ISCAS-85 benchmark suite, comprising $11$ combinational circuits \cite{Hansen1999ISCASBench}. The results of {\sc Demotic} were obtained from running on a system equipped with an Intel Xeon E$5-2698$ with $2.2$GHz clock rate and $8$ NVIDIA V$100$ GPUs with $32$GB of memory each. We report the runtime performance of {\sc Demotic} in term of throughput, measured as the number of valid and distinct solutions per second, while using a single NVIDIA V$100$ GPU. To obtain the experimental results of {\sc Demotic} for the CircuitSAT problems of the ISCAS-85 benchmark suite, we used GD as the optimizer. We set the learning rate to $15$, the batch size to $500,000$, and the number of iterations to $10$.

% , and the ISCAS-89 benchmark suite, comprising $47$ sequential circuits

\subsection{Runtime Performance}
We use all $11$ combinational circuits from the ISCAS-85 benchmark suite, encompassing designs ranging from relatively simple to moderately complex. These circuits serve as standardized test cases for evaluating algorithm performance in tasks such as logic synthesis, technology mapping, simulation, and testing. We convert these circuits into CircuitSAT sampling problems by randomly assigning specific binary values to some of their output nodes. The objective is to identify a set of inputs that would yield the desired values for those fixed outputs. The size of the solution space for such problems is proportional to the number of inputs in these circuits. Table \ref{tab2} summarizes the sampling performance of {\sc Demotic} in terms of throughput for all the combinational circuits in the ISCAS-85 benchmark suite. Throughput is measured as the number of unique solutions generated per second. We report the experimental results corresponding to the best throughput obtained from each sampler in Table \ref{tab2}.

\begin{table*}[t]
    \centering
    \caption{The runtime performance of {\sc Demotic}, {\sc UniGen3}, {\sc CMSGen} and {\sc DiffSampler} is evaluated in terms of unique solution throughput. Throughput is measured under the case where each method is tasked with generating a minimum of $1000$ distinct solutions (except for ``c17'') within a timeout (TO) of $2$ hours.}
    \vspace{-0.25cm}
    \include{table2}

    \label{tab2}
    \vspace{-0.25cm}
\end{table*}

For comparison purposes, we evaluate {\sc Demotic}'s performance against state-of-the-art SAT sampler baselines, namely {\sc UniGen3} \cite{Soos2020unigen3}, {\sc CMSGen} \cite{Golia2021cmsgen}, and {\sc DiffSampler} \cite{Ardakani2024diffsampler}. {\sc UniGen3} and {\sc CMSGen} are highly optimized C++ implementations, whereas {\sc DiffSampler} is a GPU-accelerated SAT sampler implemented in Python using the high-performance numerical computing library JAX. To this end, we first need to convert the CircuitSAT problems into their CNF formulas under the same aforementioned output constraints. We employ the Tseytin transformation, which takes a combinational logic circuit as input and produces its corresponding CNF \cite{tseitin1983complexity}. The size of the solution space for the resulting SAT problems is proportional to the number of variables in their CNF representation. Table \ref{tab2} presents the performance of the baseline samplers for the obtained SAT instances. {\sc UniGen3} and {\sc CMSGen} were executed on server-grade Intel Xeon Gold $6254$ CPU with a clock rate of $3.1$GHz and $790$GB of RAM. Similar to {\sc Demotic}, the results of {\sc DiffSampler} were obtained from running on a system equipped with an Intel Xeon E$5-2698$ with $2.2$GHz clock rate and $8$ NVIDIA V$100$ GPUs with $32$GB of memory each.

% \begin{table*}[t]
%     \centering
%     \caption{The runtime performance of {\sc Demotic} is evaluated in terms of unique solution throughput for sequential circuits.}
%     \include{table3}
%     \label{tab3}
%     \vspace{-0.25cm}
% \end{table*}

The experimental results presented in Table \ref{tab2} showcase the superior performance of {\sc Demotic} in the sampling task, surpassing state-of-the-art samplers by over two orders of magnitude in most cases. This is because the conversion to CNF introduces additional variables and operations in the form of clauses depending on the complexity of the underlying circuit, as shown in Table \ref{tab2}, undermining the performance of baseline samplers across all the CircuitSAT instances except for ``c$17$''. Due to the limited number of inputs in the CircuitSAT instance for ``c$17$'', only $18$ unique solutions exist when constraining the circuit's second output to $1$. This restriction reduces {\sc Demotic}'s performance in this scenario, as the GPU becomes under-utilized. Consequently, {\sc CMSGen} performs more efficiently in this case.

\begin{figure}
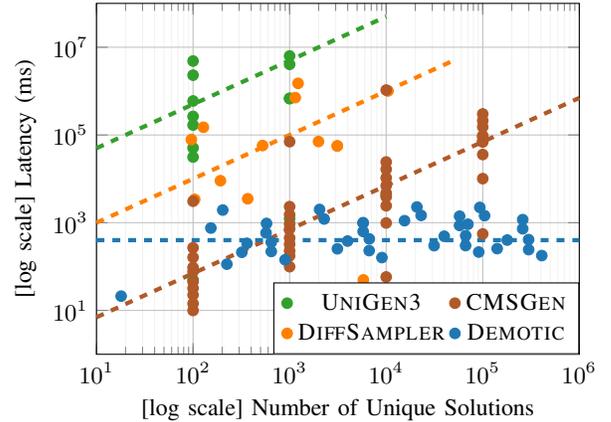

    \centering
    \include{figure3}
    \vspace{-1cm}
    \caption{Log-Log plot of runtime in millisecond against the
count of unique satisfying solutions found within that run time across all the CircuitSAT problems from the ISCAS-85 benchmark. Dotted lines denote the trend of the runtime performance for each sampler.}
    \label{fig3}
    \vspace{-0.5cm}
\end{figure}

Fig. \ref{fig3} illustrates the scaling patterns of runtime performance relative to the number of unique solutions generated by each sampler. The analysis reveals two key findings: Firstly, {\sc Demotic} overall demonstrates superior efficiency compared to {\sc UniGen3}, {\sc CMSGen} and {\sc DiffSampler}, especially when sampling larger numbers of solutions. Secondly, our method exhibits more efficient scalability, as evidenced by the linear scaling of the time required for generating larger numbers of solutions.

\vspace{-0.25cm}
\subsection{Learning Dynamics}
In this section, we provide a detailed analysis of the learning dynamics of {\sc Demotic} over time. For all experiments, we set the batch size to $500,000$ and the learning rate to $15$ unless stated otherwise. We excluded the module ``c$17$'' from our experiments due to its limited number of inputs.

\begin{figure}[t]
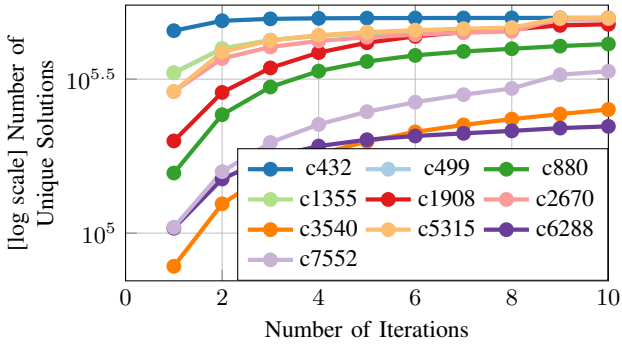

    \centering
    \vspace{-0.75cm}
    \include{figure4}
    \vspace{-1cm}
    \caption{Log learning plot of {\sc Demotic} showing the number of unique satisfying solutions across different iterations for the CircuitSAT problems from the ISCAS-85 benchmark.}
    \label{fig4}
    \vspace{-0.25cm}
\end{figure}

Figure \ref{fig4} illustrates the learning progress of {\sc Demotic} in terms of the number of unique solutions discovered across $10$ iterations. The learning curves show that as the number of iterations increases, the number of unique solutions learned by {\sc Demotic} also increases. While there is no theoretical guarantee that gradient descent will reach the global minimum in non-convex landscapes, including our continuous formulation of CircuitSAT problems, our experiments demonstrate its effectiveness in finding solutions that perform well in the continuous form. Even if these solutions aren't the absolute global minimum in the continuous form, they still satisfy the CircuitSAT constraints in the discrete form.

\begin{figure}[t]
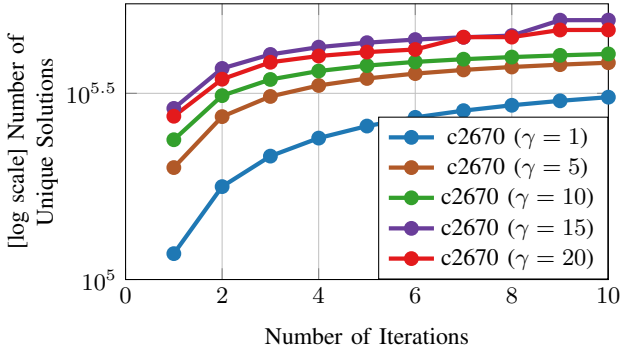

    \centering
    \vspace{-0.5cm}
    \include{figure6}
    \vspace{-1cm}
    \caption{Log learning plot of {\sc Demotic} showing the number of unique satisfying solutions across different iterations and learning rates for the CircuitSAT problem of ``c$2670$''.}
    \label{fig6}
    \vspace{-0.5cm}
\end{figure}

The convergence rate for each CircuitSAT problem varies depending on the complexity and structure of the underlying circuit, as well as the chosen hyper-parameters, such as the learning rate. More complex circuits and sub-optimal hyper-parameter settings typically result in slower convergence rates. Conversely, simpler circuits and well-tuned hyper-parameters tend to lead to faster convergence. Choosing an appropriate learning rate, as the most important hyper-parameter in our experiments, is crucial for effective model training. Very low learning rates can make slow convergence of the learning process, requiring many iterations to reach an optimal solution, which increases computational cost and time. On the other hand, very high learning rates can make the model oscillating around the minimum and leading to poor convergence. For instance, Fig. \ref{fig6} illustrates the learning progress of {\sc Demotic} across different learning rates ranging from $1$ to $20$ for the CircuitSAT problem ``c$2670$''. In this example, the learning rate of $15$ provides the best convergence among the tested rates. In contrast, the learning rate of $1$ results in the slowest convergence, while the learning rate of $20$ leads to slower convergence than the learning rate of $15$.

\begin{figure}[t]
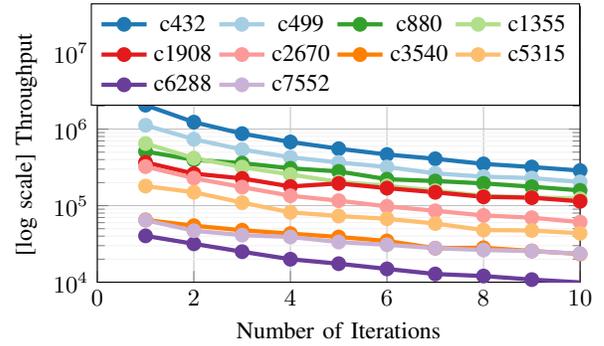

    \centering
    \include{figure5}
    \vspace{-1cm}
    \caption{Log plot of the throughput of {\sc Demotic}, measured by the number of unique satisfying solutions per second across different iterations for the CircuitSAT problems from the ISCAS-85 benchmark.}
    \label{fig5}
    \vspace{-0.25cm}
\end{figure}

While increasing the number of iterations can lead to learning more unique solutions, it does not necessarily result in higher throughput, as shown in Fig. \ref{fig5}. In fact, the majority of solutions are learned by the end of the first iteration. Specifically, the number of solutions at the end of the first iteration is higher than the number of new unique solutions learned at the end of each subsequent iteration. Given that the latency of each iteration is roughly the same, the throughput of generating unique solutions decreases as the number of iterations increases, as depicted in Fig. \ref{fig5}.

\begin{figure}[t]
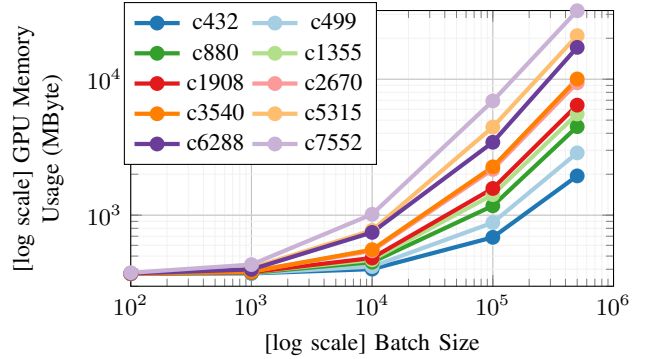

    \centering
    \vspace{-0.5cm}
    \include{figure7}
    \vspace{-1cm}
    \caption{Log-log plot of GPU memory usage of {\sc Demotic} in megabytes, measured by ``nvidia-smi'' across different batch sizes for the CircuitSAT problems from the ISCAS-85 benchmark.}
    \label{fig7}
    \vspace{-0.5cm}
\end{figure}

This observation suggests that running {\sc Demotic} for only one iteration may be sufficient to obtain the desired number of distinct solutions by adjusting the batch size. However, this conclusion holds only when there is no GPU memory constraint for the underlying circuit. GPU acceleration of CircuitSAT sampling incurs GPU memory usage depending on the size of the CircuitSAT problem and the batch size. Fig. \ref{fig7} shows the GPU memory usage of the CircuitSAT problems, measured by ``nvidia-smi'', across different batch sizes. This figure illustrates the significant growth in GPU memory usage for larger batch sizes. In scenarios where generating a large number of unique samples is targeted but there are constraints on GPU memory usage, the inevitable solution is to run the learning process for more iterations, albeit at the cost of lower throughput.

% \subsection{Sequential Circuits}
% In the task of sequential circuit sampling, we employ the entire ISCAS-89 benchmark suite, comprising all its components. Unlike in combinational circuit sampling, the objective here is to discover input sequences that fulfill the specified output constraints for a certain number of clock cycles. To transform these circuits into CircuitSAT sampling problems, we randomly assign specific binary values to some of their output nodes after a certain number of clock cycles. We also set the number clock cycles to $25$ for these experiments. In fact, we treat the number of clock cycles as a hyper-parameter for the sampling task. Table \ref{tab3} summarizes the sampling performance of {\sc Demotic} for sequential circuits in the ISCAS-89 benchmark suite. Due to the temporal nature of sequential circuits, the sampling task is more challenging compared to the combinational circuits. As such, the sampling throughput of {\sc Demotic} for sequential circuits is lower than that for combinational circuits. It is worth mentioning that we report the experimental results of {\sc Demotic} for sequential circuits, as the conversion of such circuits to CNF is non-trivial, as discussed in Section \ref{subsec:circuitsat_sampling}. 

%% file: table2.tex
\begin{tabular}{c|c|c|c|c|c|c|c|c|c} 
\hline
\midrule
CircuitSAT  & \multirow{ 2}{*}{\# Inputs} & \multirow{ 2}{*}{\# Outputs} & \# Logic & \# Variables & \# Clauses & \multirow{ 2}{*}{{\sc Demotic}} & \multirow{ 2}{*}{{\sc UniGen3}} & \multirow{ 2}{*}{{\sc CMSGen}} & \multirow{ 2}{*}{{\sc DiffSampler}}\\ 
Instance &  &  & Gates & (CNF) &  (CNF) & &  & &\\ 
 \midrule
c$17$    & $5$   & $2$   & $6$    & $25$   & $19$   & $850$       & $15$ & $\mathbf{2,928}$ & 36\\
c$432$   & $36$  & $7$   & $160$  & $539$  & $516$  & $\mathbf{2,054,518}$ & $1.5$ & $10,070$ & 105 \\
c$499$   & $41$  & $32$  & $202$  & $683$  & $717$  & $\mathbf{1,123,605}$ & $1.5$ & $5,704$ & 28\\
c$880$   & $60$  & $26$  & $383$  & $1198$ & $1115$ & $\mathbf{510,760}$   & $0.2$ & $4,379$ & 15\\
c$1355$  & $41$  & $32$  & $546$  & $1683$ & $1613$ & $\mathbf{648,736}$   & $0.2$ & $3,109$ & 0.9 \\
c$1908$  & $33$  & $25$  & $880$  & $2436$ & $2381$ & $\mathbf{367,720}$   & TO & $2,213$ & TO\\
c$2670$  & $233$ & $140$ & $1269$ & $3642$ & $3274$ & $\mathbf{323,617}$   & TO & $1,385$ & TO\\
c$3540$  & $50$  & $22$  & $1669$ & $4680$ & $4611$ & $\mathbf{65,156}$    & TO & $1,073$ & TO\\
c$5315$  & $178$ & $123$ & $2307$ & $6994$ & $6698$ & $\mathbf{180,085}$   & TO & $655$ & TO\\
c$6288$  & $32$  & $32$  & $2416$ & $7280$ & $7219$ & $\mathbf{40,325}$    & TO & $14$ & TO\\
c$7552$  & $207$ & $108$ & $3513$ & $9971$ & $9661$ & $\mathbf{64,483}$    & TO & $430$ & TO\\
\midrule
\hline
\end{tabular}

%% file: figure3.tex
\begin{tikzpicture}
    \begin{axis}[
        xlabel={[log scale] Number of Unique Solutions},
        ylabel={[log scale] Latency (ms)},
        xmin=10, xmax=1000000,
        ymin=1, ymax=100000000,
        xmode=log,
        ymode=log,
        grid=both,
        grid style={line width=.1pt, draw=gray!10},
        major grid style={line width=.2pt,draw=gray!50},
        xlabel near ticks,
        ylabel near ticks,
        tick align=inside,
        legend columns=2,
        transpose legend,
        height=6.25cm, width=8cm,
        legend style={at={(1,0)},anchor=south east},
    ]
    \addplot[only marks, mark=*, color=Paired-3] coordinates {
        (100, 50)
        (100, 31380)
        (100, 50840)
        (100, 168670)
        (100, 265120)
        (100, 595610)
        (100, 2309270)
        (100, 4850690)
        (1000, 1230)
        (1000, 687550)
        (1000, 671140)
        (1000, 4068700)
        (1000, 6326400)
        %(10000, 73820)
    };
    \addplot[ultra thick,domain=10:10000, color=Paired-3, dashed, forget plot]{5000*x};
    \addlegendentry{\sc UniGen3}

    \addplot[only marks, mark=*, color=Paired-7] coordinates {
        % (7528, 248.951)
        (5806, 49.6365)
        % (2457, 35.67791)
        % (2057, 18.6316)
        % (20832, 292.757)
        % (57295, 503.769)
        % (7415, 249.47)
        (368, 3516.54)
        (104, 3399.64)
        (525, 57454.167)
        (3118, 56250.77)
        (192, 9059.71)
        (1992, 71016.24)
        (1144, 710450.22)
        (10438, 1007498.608)
        (96, 78667.00)
        (1224, 1503129.80)
        (128, 149923.86)
        (3118, 56250.077)
        % (63805, 972.33)
    };
    \addplot[ultra thick,domain=10:50000, color=Paired-7, dashed, forget plot]{100*x};
    \addlegendentry{\sc DiffSampler}

    \addplot[only marks, mark=*, color=Paired-11] coordinates {
        (100, 0.957)
        (100, 9.97)
        (100, 14.2943)
        (100, 22.5953)
        (100, 32.0733)
        (100, 45.3653)
        (100, 76.7366)
        (100, 95.4796)
        (100, 162.196)
        (100, 3087.6266)
        (100, 268.6063)
        
        (1000, 6.146)
        (1000, 99.304)
        (1000, 175.316)
        (1000, 228.353)
        (1000, 321.679)
        (1000, 451.804)
        (1000, 721.849)
        (1000, 932.271)
        (1000, 1525.992)
        (1000, 70386.504)
        (1000, 2324.909)

        (10000, 57.9435)
        (10000, 984.886)
        (10000, 3883.8815)
        (10000, 2428.1915)
        (10000, 4226.684)
        (10000, 4748.4845)
        (10000, 6971.69)
        (10000, 10535.913)
        (10000, 16559.4415)
        (10000, 1056347.715)
        (10000, 24240.393)

        (100000, 553.655)
        (100000, 10103.162)
        (100000, 91483.272)
        (100000, 35831.97)
        (100000, 96164.261)
        (100000, 69139.546)
        (100000, 74110.951)
        (100000, 152873.802)
        (100000, 212467.212)
        (100000, 301737.367)
    };
    \addplot[ultra thick,domain=10:1000000, color=Paired-11, dashed, forget plot]{0.7*x};
    \addlegendentry{\sc CMSGen}

    \addplot[only marks, mark=*, color=Paired-1] coordinates {
        (18, 21.173)
        
        (898, 142.85)
        (646, 220.697)
        (320, 211.817)
        (646, 0358.768)
        (362, 338.079)
        (570, 585.66)
        (153, 754.719)
        (577, 961.079)
        (203, 1940.79)
        (224, 113.416)

        (9040, 160.339)
        (6671, 231.103)
        (3125, 254.342)
        (6688, 427.362)
        (3983, 381.682)
        (5776, 630.676)
        (5770, 994.876)
        (5770, 988.922)
        (2043, 2003.499)
        (2280, 1214.748)

        (90885, 214.515)
        (66481, 300.639)
        (31355, 299.246)
        (66619, 507.591)
        (39928, 490.695)
        (57557, 864.46)
        (15486, 1111.376)
        (57611, 1407.184)
        (20812, 2282.095)
        (23010, 1467.15)
        
        (408763, 177.944)
        (298997, 243.306)
        (141098, 256.507)
        (298621, 411.157)
        (179406, 401.845)
        (259258, 726.907)
        (70263, 922.172)
        (259493, 1173.812)
        (93242, 2231.983)
        (104476, 1448.933)
    };
    \addlegendentry{\sc Demotic}
    \addplot[ultra thick,domain=10:1000000, color=Paired-1, dashed, forget plot]{0.000 * x+400}; % +400
    
    \end{axis}
\end{tikzpicture}

%% file: figure4.tex
\begin{tikzpicture}
    \begin{axis}[
        xlabel={Number of Iterations},
        ylabel={[log scale] Number of \\ Unique Solutions},
        ylabel style={align=center, text width=5cm},
        xmin=0, xmax=10,
        ymin=70000, ymax=550000,
        ymode=log,
        grid=both,
        grid style={line width=.1pt, draw=gray!10},
        major grid style={line width=.2pt,draw=gray!50},
        xlabel near ticks,
        % ylabel near ticks,
        tick align=inside,
        legend columns=3,
        height=5.25cm, width=8cm,
        % transpose legend,
        legend style={at={(1,0)},anchor=south east},
    ]
    \addplot[ultra thick, mark=*, color=Paired-1] coordinates {
        (1, 454217)
        (2, 489166)
        (3, 496276)
        (4, 498272)
        (5, 499038)
        (6, 499357)
        (7, 499559)
        (8, 499660)
        (9, 499744)
        (10, 499801)
    };
    \addlegendentry{c432}

    \addplot[ultra thick, mark=*, color=Paired-2] coordinates {
        (1, 331213)
        (2, 397056)
        (3, 423272)
        (4, 437596)
        (5, 446734)
        (6, 453455)
        (7, 458512)
        (8, 462589)
        (9, 489046)
        (10, 492910)
    };
    \addlegendentry{c499}
    
    \addplot[ultra thick, mark=*, color=Paired-3] coordinates {
        (1, 156693)
        (2, 242410)
        (3, 298202)
        (4, 335736)
        (5, 360808)
        (6, 377665)
        (7, 388961)
        (8, 396845)
        (9, 404904)
        (10, 411186)
    };
    \addlegendentry{c880}
    
    \addplot[ultra thick, mark=*, color=Paired-4] coordinates {
        (1, 331905)
        (2, 397424)
        (3, 423202)
        (4, 437169)
        (5, 445826)
        (6, 452066)
        (7, 456572)
        (8, 460130)
        (9, 493596)
        (10, 493679)
    };
    \addlegendentry{c1355}
    
    \addplot[ultra thick, mark=*, color=Paired-5] coordinates {
        (1, 199217)
        (2, 286288)
        (3, 343927)
        (4, 385617)
        (5, 415136)
        (6, 435375)
        (7, 449277)
        (8, 458513)
        (9, 472642)
        (10, 476544)
    };
    \addlegendentry{c1908}

    \addplot[ultra thick, mark=*, color=Paired-6] coordinates {
        (1, 288101)
        (2, 368871)
        (3, 402149)
        (4, 420736)
        (5, 432469)
        (6, 440893)
        (7, 447143)
        (8, 451833)
        (9, 497034)
        (10, 497234)
    };
    \addlegendentry{c2670}

    \addplot[ultra thick, mark=*, color=Paired-7] coordinates {
        (1, 78029)
        (2, 124475)
        (3, 156935)
        (4, 180468)
        (5, 198377)
        (6, 213104)
        (7, 224537)
        (8, 234566)
        (9, 243503)
        (10, 251998)
    };
    \addlegendentry{c3540}

    \addplot[ultra thick, mark=*, color=Paired-8] coordinates {
        (1, 288387)
        (2, 387267)
        (3, 422490)
        (4, 439356)
        (5, 449264)
        (6, 456067)
        (7, 460995)
        (8, 464695)
        (9, 499184)
        (10, 499414)
    };
    \addlegendentry{c5315}

    \addplot[ultra thick, mark=*, color=Paired-9] coordinates {
        (1, 103703)
        (2, 149924)
        (3, 176449)
        (4, 191701)
        (5, 201012)
        (6, 206724)
        (7, 210844)
        (8, 214727)
        (9, 219250)
        (10, 222165)
    };
    \addlegendentry{c6288}

    \addplot[ultra thick, mark=*, color=Paired-10] coordinates {
        (1, 104476)
        (2, 158323)
        (3, 196955)
        (4, 225396)
        (5, 247804)
        (6, 266135)
        (7, 281587)
        (8, 294938)
        (9, 326810)
        (10, 335439)
    };
    \addlegendentry{c7552}

    \end{axis}
\end{tikzpicture}

%% file: figure6.tex
\begin{tikzpicture}
    \begin{axis}[
        xlabel={Number of Iterations},
        ylabel={[log scale] Number of \\ Unique Solutions},
        ylabel style={align=center, text width=5cm},
        xmin=0, xmax=10,
        ymin=100000, ymax=550000,
        ymode=log,
        grid=both,
        grid style={line width=.1pt, draw=gray!10},
        major grid style={line width=.2pt,draw=gray!50},
        % xlabel near ticks,
        % ylabel near ticks,
        tick align=inside,
        legend columns=1,
        height=5.25cm, width=8cm,
        % transpose legend,
        legend style={at={(1,0)},anchor=south east},
    ]
    \addplot[ultra thick, mark=*, color=Paired-1] coordinates {
        (1, 117369)
        (2, 177630)
        (3, 214478)
        (4, 239753)
        (5, 258289)
        (6, 272555)
        (7, 284107)
        (8, 293721)
        (9, 301874)
        (10, 308798)
    };
    \addlegendentry{c2670 ($\gamma = 1$)}

    \addplot[ultra thick, mark=*, color=Paired-11] coordinates {
        (1, 199856)
        (2, 273810)
        (3, 310177)
        (4, 331970)
        (5, 346636)
        (6, 357269)
        (7, 365502)
        (8, 372154)
        (9, 377588)
        (10, 382180)
    };
    \addlegendentry{c2670 ($\gamma = 5$)}

    \addplot[ultra thick, mark=*, color=Paired-3] coordinates {
        (1, 237342)
        (2, 311749)
        (3, 344527)
        (4, 363006)
        (5, 375423)
        (6, 384051)
        (7, 390512)
        (8, 395716)
        (9, 399956)
        (10, 403410)
    };
    \addlegendentry{c2670 ($\gamma = 10$)}

    \addplot[ultra thick, mark=*, color=Paired-9] coordinates {
        (1, 288101)
        (2, 368871)
        (3, 402149)
        (4, 420736)
        (5, 432469)
        (6, 440893)
        (7, 447143)
        (8, 451833)
        (9, 497034)
        (10, 497234)
    };
    \addlegendentry{c2670 ($\gamma = 15$)}

    \addplot[ultra thick, mark=*, color=Paired-5] coordinates {
        (1, 274447)
        (2, 345123)
        (3, 383208)
        (4, 398351)
        (5, 408001)
        (6, 414634)
        (7, 447435)
        (8, 447639)
        (9, 467791)
        (10, 467936)
    };
    \addlegendentry{c2670 ($\gamma = 20$)}

    \end{axis}
\end{tikzpicture}

%% file: figure5.tex
\begin{tikzpicture}
    \begin{axis}[
        xlabel={Number of Iterations},
        ylabel={[log scale] Throughput},
        xmin=0, xmax=10,
        ymin=10000, ymax=40000000,
        ymode=log,
        grid=both,
        grid style={line width=.1pt, draw=gray!10},
        major grid style={line width=.2pt,draw=gray!50},
        xlabel near ticks,
        ylabel near ticks,
        tick align=inside,
        legend columns=4,
        height=5.25cm, width=8cm,
        % transpose legend,
        legend style={at={(1,0.64)},anchor=south east},
    ]
    \addplot[ultra thick, mark=*, color=Paired-1] coordinates {
        (1, 454217/0.221082)
        (2, 489166/0.396940)
        (3, 496276/0.569149)
        (4, 498272/0.733313)
        (5, 499038/0.899716)
        (6, 499357/1.073187)
        (7, 499559/1.219987)
        (8, 499660/1.423881)
        (9, 499744/1.562670)
        (10, 499801/1.740273)
    };
    \addlegendentry{c432}

    \addplot[ultra thick, mark=*, color=Paired-2] coordinates {
        (1, 331213/0.294777)
        (2, 397056/0.538940)
        (3, 423272/0.783097)
        (4, 437596/1.021012)
        (5, 446734/1.21888)
        (6, 453455/1.417753)
        (7, 458512/1.73467)
        (8, 462589/1.933012)
        (9, 489046/2.133596)
        (10, 492910/2.411745)
    };
    \addlegendentry{c499}
    
    \addplot[ultra thick, mark=*, color=Paired-3] coordinates {
        (1, 156693/0.306784)
        (2, 242410/0.611079)
        (3, 298202/0.829638)
        (4, 335736/1.09174)
        (5, 360808/1.288729)
        (6, 377665/1.704616)
        (7, 388961/1.853998)
        (8, 396845/2.030747)
        (9, 404904/2.291631)
        (10, 411186/2.597413)
    };
    \addlegendentry{c880}
    
    \addplot[ultra thick, mark=*, color=Paired-4] coordinates {
        (1, 331905/0.511618)
        (2, 397424/0.950628)
        (3, 423202/1.328398)
        (4, 437169/1.712766)
        (5, 445826/2.183245)
        (6, 452066/2.48913)
        (7, 456572/2.912655)
        (8, 460130/3.431903)
        (9, 493596/3.828706)
        (10, 493679/4.019502)
    };
    \addlegendentry{c1355}
    
    \addplot[ultra thick, mark=*, color=Paired-5] coordinates {
        (1, 199217/0.541762)
        (2, 286288/1.09299)
        (3, 343927/1.516026)
        (4, 385617/2.16696)
        (5, 415136/2.12745)
        (6, 435375/2.57053)
        (7, 449277/3.008229)
        (8, 458513/3.519824)
        (9, 472642/3.721308)
        (10, 476544/4.18253)
    };
    \addlegendentry{c1908}

    \addplot[ultra thick, mark=*, color=Paired-6] coordinates {
        (1, 288101/0.890251)
        (2, 368871/1.612641)
        (3, 402149/2.304325)
        (4, 420736/3.122802)
        (5, 432469/3.722599)
        (6, 440893/4.507569)
        (7, 447143/5.217566)
        (8, 451833/6.06830)
        (9, 497034/7.142945)
        (10, 497234/8.1555)
    };
    \addlegendentry{c2670}

    \addplot[ultra thick, mark=*, color=Paired-7] coordinates {
        (1, 78029/1.19757)
        (2, 124475/2.276511)
        (3, 156935/3.28948)
        (4, 180468/4.18386)
        (5, 198377/5.08798)
        (6, 213104/6.16538)
        (7, 224537/8.07635)
        (8, 234566/8.3279)
        (9, 243503/9.510849)
        (10, 251998/10.792431)
    };
    \addlegendentry{c3540}

    \addplot[ultra thick, mark=*, color=Paired-8] coordinates {
        (1, 288387/1.60139)
        (2, 387267/2.587435)
        (3, 422490/3.855495)
        (4, 439356/5.370636)
        (5, 449264/6.14065)
        (6, 456067/6.770219)
        (7, 460995/7.899824)
        (8, 464695/9.687201)
        (9, 499184/10.537846)
        (10, 499414/11.473759)
    };
    \addlegendentry{c5315}

    \addplot[ultra thick, mark=*, color=Paired-9] coordinates {
        (1, 103703/2.57166)
        (2, 149924/4.723804)
        (3, 176449/7.05026)
        (4, 191701/9.60294)
        (5, 201012/11.523619)
        (6, 206724/13.86723)
        (7, 210844/16.48168)
        (8, 214727/17.83167)
        (9, 219250/20.28465)
        (10, 222165/22.338234)
    };
    \addlegendentry{c6288}

    \addplot[ultra thick, mark=*, color=Paired-10] coordinates {
        (1, 104476/1.620207)
        (2, 158323/3.366065)
        (3, 196955/4.783319)
        (4, 225396/5.780267)
        (5, 247804/7.40293)
        (6, 266135/8.630015)
        (7, 281587/10.125299)
        (8, 294938/11.199752)
        (9, 326810/12.826562)
        (10, 335439/14.246239)
    };
    \addlegendentry{c7552}

    \end{axis}
\end{tikzpicture}

%% file: figure7.tex
\begin{tikzpicture}
    \begin{axis}[
        xlabel={[log scale] Batch Size},
        ylabel={[log scale] GPU Memory \\ Usage (MByte)},
        ylabel style={align=center, text width=5cm},
        xmin=100, xmax=1000000,
        ymin=300, ymax=32000,
        xmode=log,
        ymode=log,
        grid=both,
        grid style={line width=.1pt, draw=gray!10},
        major grid style={line width=.2pt,draw=gray!50},
        % xlabel near ticks,
        % ylabel near ticks,
        tick align=inside,
        legend columns=2,
        height=5.25cm, width=8cm,
        legend style={at={(0.51,0.44)},anchor=south east},
    ]
    \addplot[ultra thick, mark=*, color=Paired-1] coordinates {
        (100, 371)
        (1000, 373)
        (10000, 401)
        (100000, 687)
        (500000, 1945)
    };
    \addlegendentry{c432}

    \addplot[ultra thick, mark=*, color=Paired-2] coordinates {
        (100, 371)
        (1000, 375)
        (10000, 419)
        (100000, 881)
        (500000, 2871)
    };
    \addlegendentry{c499}
    
    \addplot[ultra thick, mark=*, color=Paired-3] coordinates {
        (100, 371)
        (1000, 377)
        (10000, 451)
        (100000, 1172)
        (500000, 4479)
    };
    \addlegendentry{c880}
    
    \addplot[ultra thick, mark=*, color=Paired-4] coordinates {
        (100, 371)
        (1000, 381)
        (10000, 469)
        (100000, 1416)
        (500000, 5551)
    };
    \addlegendentry{c1355}
    
    \addplot[ultra thick, mark=*, color=Paired-5] coordinates {
        (100, 371)
        (1000, 381)
        (10000, 484)
        (100000, 1572)
        (500000, 6447)
    };
    \addlegendentry{c1908}

    \addplot[ultra thick, mark=*, color=Paired-6] coordinates {
        (100, 371)
        (1000, 387)
        (10000, 547)
        (100000, 2172)
        (500000, 9411)
    };
    \addlegendentry{c2670}

    \addplot[ultra thick, mark=*, color=Paired-7] coordinates {
        (100, 371)
        (1000, 384)
        (10000, 557)
        (100000, 2266)
        (500000, 10043)
    };
    \addlegendentry{c3540}

    \addplot[ultra thick, mark=*, color=Paired-8] coordinates {
        (100, 375)
        (1000, 409)
        (10000, 773)
        (100000, 4447)
        (500000, 20925)
    };
    \addlegendentry{c5315}

    \addplot[ultra thick, mark=*, color=Paired-9] coordinates {
        (100, 373)
        (1000, 399)
        (10000, 747)
        (100000, 3437)
        (500000, 17148)
    };
    \addlegendentry{c6288}

    \addplot[ultra thick, mark=*, color=Paired-10] coordinates {
        (100, 377)
        (1000, 433)
        (10000, 1015)
        (100000, 6920)
        (500000, 31880)
    };
    \addlegendentry{c7552}

    \end{axis}
\end{tikzpicture}

%% file: 02_background.tex
\section{Related Work}
Several SAT formula sampling techniques have been explored in the literature. {\sc UniGen3}, for instance, offers approximate uniformity guarantees \cite{yash2022barbarik}, while {\sc CMSGen} and {\sc Quicksampler} \cite{dutra2018quicksampler} prioritize sampling efficiency. Previous research has also investigated the use of data-parallel hardware for SAT solving, primarily focusing on parallelizing CDCL or other heuristic-based SAT solving algorithms \cite{costa2013parallelization, osama2021sat}. Attempts have been made to frame a SAT instance as a constrained numerical optimization problem, as seen in recent work like {\sc MatSat} \cite{sato2021matsat} and {\sc NeuroSAT} \cite{amizadeh2018learning}. Nevertheless, these approaches have fallen short in showcasing the efficacy of GPU-accelerated formula sampling on standard benchmarks, which are larger and more diverse than the small, random instances typically examined in earlier research. A new differentiable sampling method named {\sc DiffSampler} was recently introduced in \cite{Ardakani2024diffsampler}. This method allows for GPU-accelerated SAT sampling on standard benchmarks and achieved a comparable runtime performance with respect to {\sc UniGen3} and {\sc CMSGen}.

%% file: 06_conclusion.tex
\vspace{-0.25cm}
\section{Conclusion}
CircuitSAT problems are typically transformed into Boolean Satisfiability (SAT) problems, where the sampling task is performed using SAT samplers, albeit with computational complexities, especially for large circuits. To reduce the computational complexity of the CircuitSAT sampling task and leverage GPU acceleration, this paper introduced a novel differentiable sampler called {\sc Demotic}. {\sc Demotic} re-frames the CircuitSAT problem as a multi-output regression task, utilizing gradient descent for learning diverse solutions. By maintaining the circuit's structure without Boolean conversion, {\sc Demotic} enables parallel learning and GPU-accelerated sampling, offering significant advancements in CircuitSAT sampling methodology. We have demonstrated the exceptional performance of {\sc Demotic} in the sampling task across various CircuitSAT instances, where it outperformed state-of-the-art samplers by more than two orders of magnitude in most cases.

%% file: main.bbl
\begin{thebibliography}{10}
\providecommand{\url}[1]{#1}
\csname url@samestyle\endcsname
\providecommand{\newblock}{\relax}
\providecommand{\bibinfo}[2]{#2}
\providecommand{\BIBentrySTDinterwordspacing}{\spaceskip=0pt\relax}
\providecommand{\BIBentryALTinterwordstretchfactor}{4}
\providecommand{\BIBentryALTinterwordspacing}{\spaceskip=\fontdimen2\font plus
\BIBentryALTinterwordstretchfactor\fontdimen3\font minus \fontdimen4\font\relax}
\providecommand{\BIBforeignlanguage}[2]{{%
\expandafter\ifx\csname l@#1\endcsname\relax
\typeout{** WARNING: IEEEtran.bst: No hyphenation pattern has been}%
\typeout{** loaded for the language `#1'. Using the pattern for}%
\typeout{** the default language instead.}%
\else
\language=\csname l@#1\endcsname
\fi
#2}}
\providecommand{\BIBdecl}{\relax}
\BIBdecl

\bibitem{Mishchenko2005Optimization}
\BIBentryALTinterwordspacing
A.~Mishchenko and R.~K. Brayton, ``Sat-based complete don't-care computation for network optimization,'' in \emph{Proceedings of the Conference on Design, Automation and Test in Europe - Volume 1}, ser. DATE '05.\hskip 1em plus 0.5em minus 0.4em\relax USA: IEEE Computer Society, 2005, p. 412–417. [Online]. Available: \url{https://doi.org/10.1109/DATE.2005.264}
\BIBentrySTDinterwordspacing

\bibitem{Tsai2009TimingAnalyzer}
\BIBentryALTinterwordspacing
S.~Tsai and C.-Y.~R. Huang, ``A false-path aware formal static timing analyzer considering simultaneous input transitions,'' in \emph{Proceedings of the 46th Annual Design Automation Conference}, ser. DAC '09.\hskip 1em plus 0.5em minus 0.4em\relax New York, NY, USA: Association for Computing Machinery, 2009, p. 25–30. [Online]. Available: \url{https://doi.org/10.1145/1629911.1629921}
\BIBentrySTDinterwordspacing

\bibitem{Bradley2011ModelChecking}
A.~R. Bradley, ``Sat-based model checking without unrolling,'' in \emph{Proceedings of the 12th International Conference on Verification, Model Checking, and Abstract Interpretation}, ser. VMCAI'11.\hskip 1em plus 0.5em minus 0.4em\relax Berlin, Heidelberg: Springer-Verlag, 2011, p. 70–87.

\bibitem{Mishchenko2006EquivalenceChecking}
\BIBentryALTinterwordspacing
A.~Mishchenko, S.~Chatterjee, R.~Brayton, and N.~Een, ``Improvements to combinational equivalence checking,'' in \emph{Proceedings of the 2006 IEEE/ACM International Conference on Computer-Aided Design}, ser. ICCAD '06.\hskip 1em plus 0.5em minus 0.4em\relax New York, NY, USA: Association for Computing Machinery, 2006, p. 836–843. [Online]. Available: \url{https://doi.org/10.1145/1233501.1233679}
\BIBentrySTDinterwordspacing

\bibitem{Zhang2021LogicSynthesis}
\BIBentryALTinterwordspacing
H.-T. Zhang, J.-H.~R. Jiang, and A.~Mishchenko, ``A circuit-based sat solver for logic synthesis,'' in \emph{2021 IEEE/ACM International Conference On Computer Aided Design (ICCAD)}.\hskip 1em plus 0.5em minus 0.4em\relax IEEE Press, 2021, p. 1–6. [Online]. Available: \url{https://doi.org/10.1109/ICCAD51958.2021.9643505}
\BIBentrySTDinterwordspacing

\bibitem{dutra2018quicksampler}
R.~Dutra, K.~Laeufer, J.~Bachrach, and K.~Sen, ``Efficient sampling of sat solutions for testing,'' in \emph{Proc. of the International Conference on Software Engineering}, 2018.

\bibitem{dutra2019EfficientSampling}
R.~T. Dutra, \emph{Efficient sampling of SAT and SMT solutions for testing and verification}.\hskip 1em plus 0.5em minus 0.4em\relax University of California, Berkeley, 2019.

\bibitem{Kitchen2007crv}
N.~Kitchen and A.~Kuehlmann, ``Stimulus generation for constrained random simulation,'' in \emph{2007 IEEE/ACM International Conference on Computer-Aided Design}, 2007, pp. 258--265.

\bibitem{Hsu2014CircuitSAT}
C.-J. Hsu, W.-H. Lin, C.-A. Wu, and K.-Y. Khoo, ``Iccad-2014 cad contest in simultaneous cnf encoder optimization with sat solver setting selection and benchmark suite,'' in \emph{Proceedings of the 2014 IEEE/ACM International Conference on Computer-Aided Design}, ser. ICCAD '14.\hskip 1em plus 0.5em minus 0.4em\relax IEEE Press, 2014, p. 357–360.

\bibitem{Velev2004CNF}
M.~N. Velev, ``Efficient translation of boolean formulas to cnf in formal verification of microprocessors,'' in \emph{Proceedings of the 2004 Asia and South Pacific Design Automation Conference}, ser. ASP-DAC '04.\hskip 1em plus 0.5em minus 0.4em\relax IEEE Press, 2004, p. 310–315.

\bibitem{Niklas2003SAT}
\BIBentryALTinterwordspacing
N.~Eén and N.~Sörensson, ``An extensible sat-solver.'' in \emph{SAT}, ser. Lecture Notes in Computer Science, E.~Giunchiglia and A.~Tacchella, Eds., vol. 2919.\hskip 1em plus 0.5em minus 0.4em\relax Springer, 2003, pp. 502--518. [Online]. Available: \url{http://dblp.uni-trier.de/db/conf/sat/sat2003.html#EenS03}
\BIBentrySTDinterwordspacing

\bibitem{Moskewicz2001Chaff}
M.~Moskewicz, C.~Madigan, Y.~Zhao, L.~Zhang, and S.~Malik, ``Chaff: engineering an efficient sat solver,'' in \emph{Proceedings of the 38th Design Automation Conference (IEEE Cat. No.01CH37232)}, 2001, pp. 530--535.

\bibitem{Audemard2018Glucose}
G.~Audemard and L.~Simon, ``On the glucose sat solver,'' \emph{International Journal on Artificial Intelligence Tools}, vol.~27, no.~01, p. 1840001, 2018.

\bibitem{Silva1996CDCL}
J.~Marques~Silva and K.~Sakallah, ``Grasp-a new search algorithm for satisfiability,'' in \emph{Proceedings of International Conference on Computer Aided Design}, 1996, pp. 220--227.

\bibitem{silva2021CDCL}
J.~Marques-Silva, I.~Lynce, and S.~Malik, \emph{\BIBforeignlanguage{English (US)}{Chapter 4: Conflict-driven clause learning SAT solvers}}, ser. Frontiers in Artificial Intelligence and Applications.\hskip 1em plus 0.5em minus 0.4em\relax IOS Press BV, 2021, pp. 133--182, publisher Copyright: {\textcopyright} 2021 The authors and IOS Press. All rights reserved.

\bibitem{Krizhevsky2012AlexNet}
\BIBentryALTinterwordspacing
A.~Krizhevsky, I.~Sutskever, and G.~E. Hinton, ``Imagenet classification with deep convolutional neural networks,'' in \emph{Advances in Neural Information Processing Systems}, F.~Pereira, C.~Burges, L.~Bottou, and K.~Weinberger, Eds., vol.~25.\hskip 1em plus 0.5em minus 0.4em\relax Curran Associates, Inc., 2012. [Online]. Available: \url{https://proceedings.neurips.cc/paper_files/paper/2012/file/c399862d3b9d6b76c8436e924a68c45b-Paper.pdf}
\BIBentrySTDinterwordspacing

\bibitem{Hansen1999ISCASBench}
M.~Hansen, H.~Yalcin, and J.~Hayes, ``Unveiling the iscas-85 benchmarks: a case study in reverse engineering,'' \emph{IEEE Design \& Test of Computers}, vol.~16, no.~3, pp. 72--80, 1999.

\bibitem{Davis1962DPLL}
\BIBentryALTinterwordspacing
M.~Davis, G.~Logemann, and D.~Loveland, ``A machine program for theorem-proving,'' \emph{Commun. ACM}, vol.~5, no.~7, p. 394–397, jul 1962. [Online]. Available: \url{https://doi.org/10.1145/368273.368557}
\BIBentrySTDinterwordspacing

\bibitem{selman1993local}
B.~Selman, H.~A. Kautz, B.~Cohen \emph{et~al.}, ``Local search strategies for satisfiability testing.'' \emph{Cliques, coloring, and satisfiability}, vol.~26, pp. 521--532, 1993.

\bibitem{Impagliazzo2017RandomSAT}
\BIBentryALTinterwordspacing
R.~Impagliazzo, V.~Kabanets, A.~Kolokolova, P.~McKenzie, and S.~Romani, ``{Does Looking Inside a Circuit Help?}'' in \emph{42nd International Symposium on Mathematical Foundations of Computer Science (MFCS 2017)}, ser. Leibniz International Proceedings in Informatics (LIPIcs), K.~G. Larsen, H.~L. Bodlaender, and J.-F. Raskin, Eds., vol.~83.\hskip 1em plus 0.5em minus 0.4em\relax Dagstuhl, Germany: Schloss Dagstuhl -- Leibniz-Zentrum f{\"u}r Informatik, 2017, pp. 1:1--1:13. [Online]. Available: \url{https://drops-dev.dagstuhl.de/entities/document/10.4230/LIPIcs.MFCS.2017.1}
\BIBentrySTDinterwordspacing

\bibitem{kitchen2009markov}
N.~Kitchen and A.~Kuehlmann, ``A markov chain monte carlo sampler for mixed boolean/integer constraints,'' in \emph{Computer Aided Verification: 21st International Conference, CAV 2009, Grenoble, France, June 26-July 2, 2009. Proceedings 21}.\hskip 1em plus 0.5em minus 0.4em\relax Springer, 2009, pp. 446--461.

\bibitem{Soos2020unigen3}
M.~Soos, S.~Gocht, and K.~S. Meel, ``Tinted, detached, and lazy cnf-xor solving and its applications to counting and sampling,'' in \emph{Proceedings of International Conference on Computer-Aided Verification (CAV)}, 2020.

\bibitem{Golia2021cmsgen}
P.~Golia, M.~Soos, S.~Chakraborty, and K.~S. Meel, ``Designing samplers is easy: The boon of testers,'' in \emph{Proc. of Formal Methods in Computer-Aided Design (FMCAD)}, 2021.

\bibitem{borchani2015survey}
H.~Borchani, G.~Varando, C.~Bielza, and P.~Larranaga, ``A survey on multi-output regression,'' \emph{Wiley Interdisciplinary Reviews: Data Mining and Knowledge Discovery}, vol.~5, no.~5, pp. 216--233, 2015.

\bibitem{Ardakani2017SC}
A.~Ardakani, F.~Leduc-Primeau, N.~Onizawa, T.~Hanyu, and W.~J. Gross, ``Vlsi implementation of deep neural network using integral stochastic computing,'' \emph{IEEE Transactions on Very Large Scale Integration (VLSI) Systems}, vol.~25, no.~10, pp. 2688--2699, 2017.

\bibitem{harris2010cmos}
D.~Harris and N.~Weste, ``Cmos vlsi design,'' \emph{ed: Pearson Education, Inc}, 2010.

\bibitem{Ardakani2024diffsampler}
A.~Ardakani, M.~Kang, K.~He, V.~Iyer, S.~Moon, and J.~Wawrzynek, ``Late breaking results: Differential and massively parallel sampling of sat formulas,'' in \emph{Proceedings of the 61st ACM/IEEE Design Automation Conference (DAC)}, 2024.

\bibitem{tseitin1983complexity}
G.~S. Tseitin, ``On the complexity of derivation in propositional calculus,'' \emph{Automation of reasoning: 2: Classical papers on computational logic 1967--1970}, pp. 466--483, 1983.

\bibitem{yash2022barbarik}
Y.~Pote and K.~S. Meel, ``On scalable testing of samplers,'' in \emph{Advances in Neural Information Processing Systems (NeurIPS)}, 2022.

\bibitem{costa2013parallelization}
C.~Costa, ``Parallelization of sat algorithms on gpus,'' Technical report, INESC-ID, Technical University of Lisbon, Tech. Rep., 2013.

\bibitem{osama2021sat}
M.~Osama, A.~Wijs, and A.~Biere, ``Sat solving with gpu accelerated inprocessing,'' in \emph{International Conference on Tools and Algorithms for the Construction and Analysis of Systems}.\hskip 1em plus 0.5em minus 0.4em\relax Springer, 2021, pp. 133--151.

\bibitem{sato2021matsat}
T.~Sato and R.~Kojima, ``Matsat: a matrix-based differentiable sat solver,'' \emph{arXiv preprint arXiv:2108.06481}, 2021.

\bibitem{amizadeh2018learning}
S.~Amizadeh, S.~Matusevych, and M.~Weimer, ``Learning to solve circuit-sat: An unsupervised differentiable approach,'' in \emph{International Conference on Learning Representations}, 2018.

\end{thebibliography}
